\documentclass[aps,prl,floats,floatfix,twocolumn,fleqn,longbibliography]{revtex4-2}

\usepackage[utf8]{inputenc}
\usepackage{dcolumn}
\usepackage{amsbsy}
\usepackage{amsmath}

\usepackage{amssymb}
\usepackage{braket}
\usepackage{color}
\usepackage{fullpage}

\usepackage{graphicx}
\graphicspath{{./}{./Figs/}{./FigsSM/}} 

\usepackage{bm}

\usepackage{braket}

\usepackage[colorlinks=true,linktoc=page,linkcolor=blue,citecolor=blue,urlcolor=blue]{hyperref}
\usepackage{cleveref}

\usepackage{epstopdf}

\newcommand{\beq}{\begin{eqnarray}}
\newcommand{\eeq}{\end{eqnarray}}
\newcommand{\bea}{\begin{eqnarray}}
\newcommand{\eea}{\end{eqnarray}}

\newcommand{\Eq}[1]{{\textcolor{blue}{Eq.}}~\!\!(\ref{#1})} 
\newcommand{\Fig}[1] {{\textcolor{blue}{Fig.}}~\!\!\ref{#1}}

\newcommand{\rmrk}[1]{#1} 
\newcommand{\cblue}[1]{{\textcolor{blue}{#1}}}

\def\<{\langle}
\def\>{\rangle}

\def\nn{\nonumber}

\renewcommand{\thesection}{\arabic{section}}

\newcommand{\sectA}[1]
{
	\addtocounter{section}{1}
	\setcounter{subsection}{0}
	\ \\
	\pdfbookmark[1]{\thesection. \ #1}{sect.\thesection}
	{\Large\bf $=\!=\!=\!=\!=\!=\;$ [\thesection] \ #1}
	\nopagebreak
	\vspace*{3mm}
}
\renewcommand{\section}{\sectA}

\newcommand{\hide}[1]{}  

\begin{document}

\title{Many-body adiabatic passage: \\ 
Instability, chaos, and quantum classical correspondence}

\author{Anant Vijay Varma$^{1,2}$,  Amichay Vardi$^{2,3}$, and Doron Cohen$^1$}
\affiliation{$^1$Department of Physics, Ben-Gurion University of the Negev, Beer Sheva, Israel}
\affiliation{$^2$Department of Chemistry, Ben-Gurion University of  the Negev, Beer Sheva, Israel}
\affiliation{$^3$ITAMP, Harvard-Smithsonian Center for Astrophysics, Cambridge, MA 02138, USA}

\begin{abstract}
Adiabatic passage in systems of interacting bosons is substantially affected by interactions and inter-particle entanglement.  We consider STIRAP-like schemes in Bose-Hubbard chains that exhibit low-dimensional chaos (a 3 site chain), and high-dimensional chaos (more than 3 sites). The dynamics that is generated by a transfer protocol exhibits striking classical and quantum chaos fingerprints that are manifested in the mean-field classical treatment, in the truncated-Wigner semiclassical treatment, and in the full many-body quantum simulations. 
\end{abstract}

\maketitle


Adiabaticity implies that a state can be followed parameterically without excitation provided the Hamiltonian is changed sufficiently slowly with respect to its characteristic frequencies \cite{Born_Fock,Kato}. This paradigm has provided some of the most prominent tools of quantum state engineering. Various adiabatic passage schemes have been utilized in order to guide systems from easily prepared initial states towards desirable target states for numerous applications in condensed matter \cite{rap_cond} and solid state physics \cite{stirap_ssp}, optics, atomic and molecular physics \cite{aps_atm,aps_mol}, optomechanics \cite{stirap_opm_1,stirap_opm_2}, quantum electrodynamics, quantum information \cite{PhysRevB.70.235317}, and chemistry\cite{stirap_1}.  

Standard adiabatic passage schemes apply to non-interacting systems that are described by single-particle theories. Notably, the Stimulated Raman Adiabatic Passage (STIRAP) \cite{stirap_1,stirap_2} scheme has been designed to transfer a particle from the first ``site" (typically an atomic eigenstate, a ro-vibronic molecular level, or a spatially localized translational state) to the 3rd site of a 3 state system.
However, the recent quantum renaissance is focused on more collective behavior in which interactions and quantum many-body entanglement play a major role.  There is thus an urgent need for better understanding of adiabatic passage in the context of many-body systems of interacting particles \cite{Ami_1,Ami_2,SM_1,amit_1,amit_2,amit_3,YW_1,YW_2,Ralf}. In particular, it is important to distinguish between interaction-induced effects that can still be captured by one-particle classical or semiclassical theories \cite{Zobay,PhysRevLett.90.170404,PhysRevA.61.023402,PhysRevA.66.023404}, and full fledged quantum effects \cite{Altland,Trimborn_2010,grafe_1,PhysRevLett.102.230401,PhysRevA.85.023605} that result from entanglement and interference between classical trajectories, and hence lead to breakdown of quantum-to-classical correspondence (QCC).

{\bf Underlying chaos.--} In this work we study the many-body dynamics during a STIRAP-type protocol designed to transfer interacting particles from the first site to the last site of an $M$ site Bose-Hubbard Chain (BHC) \cite{exprBHH1,exprBHH2,atomtronics}. 
A key observation is that \rmrk{the underlying classical phasespace is chaotic. More precisely, it is a mixed phase-space with coexistence of quasi-regular and chaotic motion} \cite{trimer2003,trimer2009,trimer2016}. Furthermore, the 3~site system features low-dimensional chaos, as opposed to larger chains that have ${d=M{-}1 > 2 }$ degrees of freedom (dof), and therefore feature high-dimensional chaos.   
\rmrk{As observed by Arnold, the difference is topological}~\cite{LLbook}:  in a 2 dof system, 2D Kolmogorov-Arnold-Moser (KAM) tori can isolate quasi-regular regions in the 3D energy surface. With more dof, the $d$-dimensional (4D for $M{=}5$) KAM tori in the $2d{-}1$ (7D for $M{=}5$) energy surface can not induce such isolation, and therefore connected chaos prevails even if the interaction is small.

{\bf Generating the dynamics.--}
We study the BHC dynamics using the following hierarchy of approaches: 
(a)~Classical mean-field theory (MFT) where a single classical trajectory is generated by the Hamiltonian; 
(b) Semiclassical truncated Wigner approximation (TWA) in which quantum uncertainty is emulated by propagating a cloud of classical trajectories; 
(c)~Quantum many-body (QMB) simulations obtained by numerically solving the time dependent Schr\"odinger equation. 
%

{\bf Outline.--} 
We introduce the BHC Hamiltonian and discuss the dependence of the transfer efficiency on the sweep rate for weak and strong interactions considering $M{=}3$ and $M{=}5$ chains. 
\rmrk{We distinguish different {\em regimes} of behavior with respect to the sweep rate, discuss the possibility of broken~QCC, and finally provide a semiclassical explanation.}      
%

\begin{figure*}
\centering

\includegraphics[width=6cm]{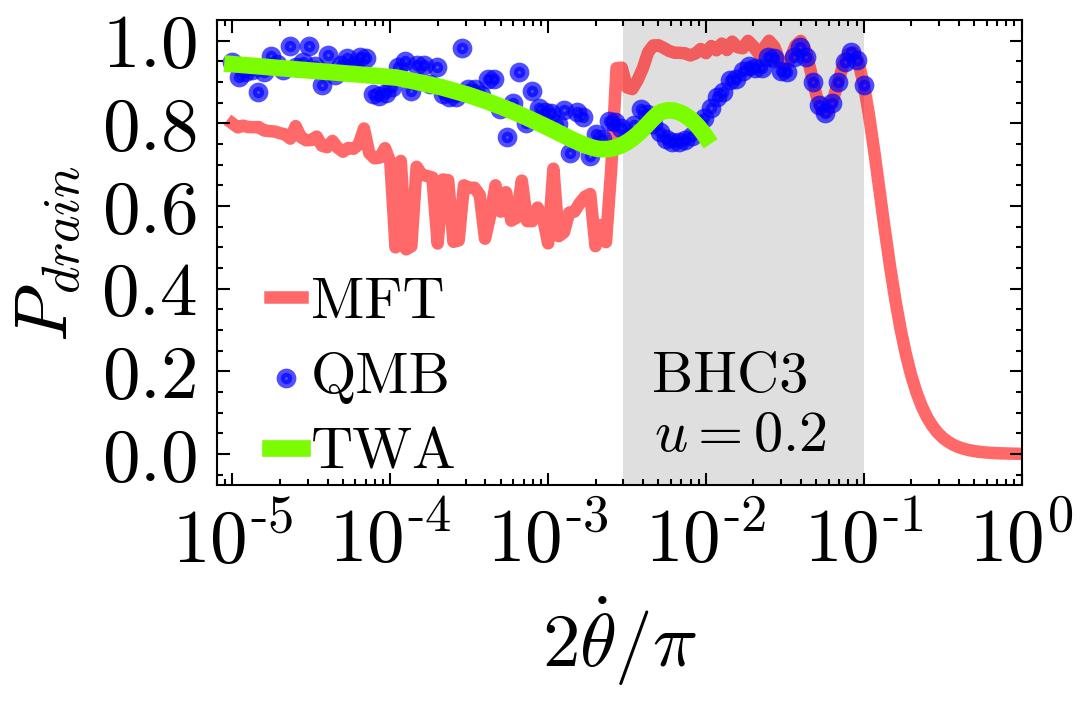}
\includegraphics[width=6cm]{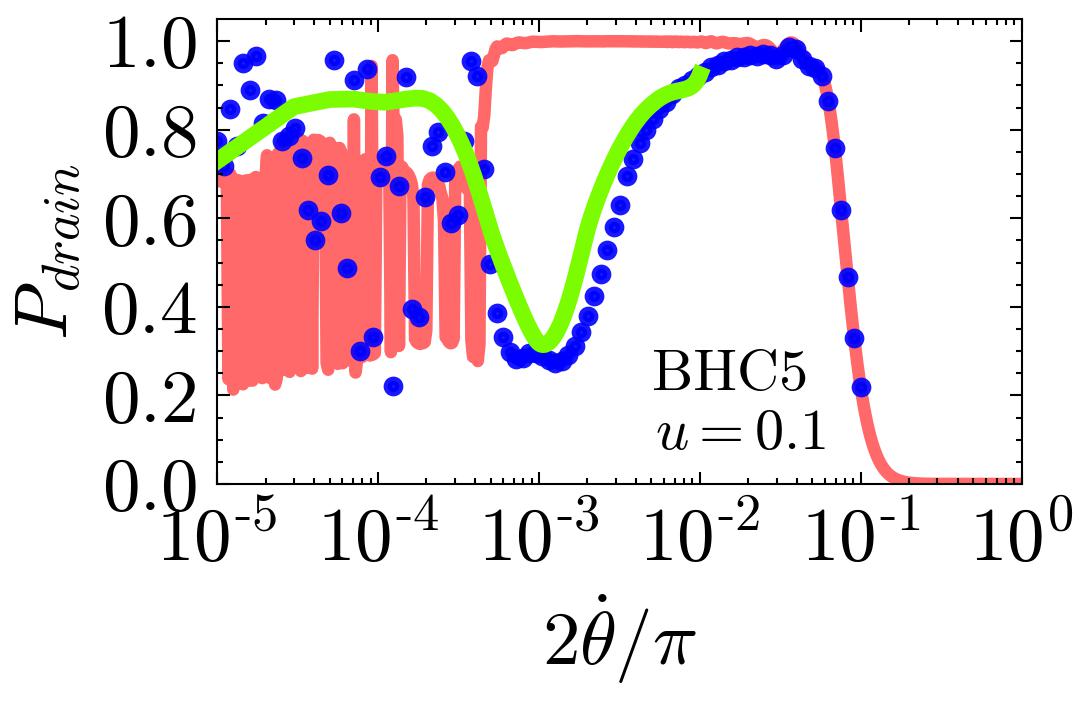}

\includegraphics[width=6cm]{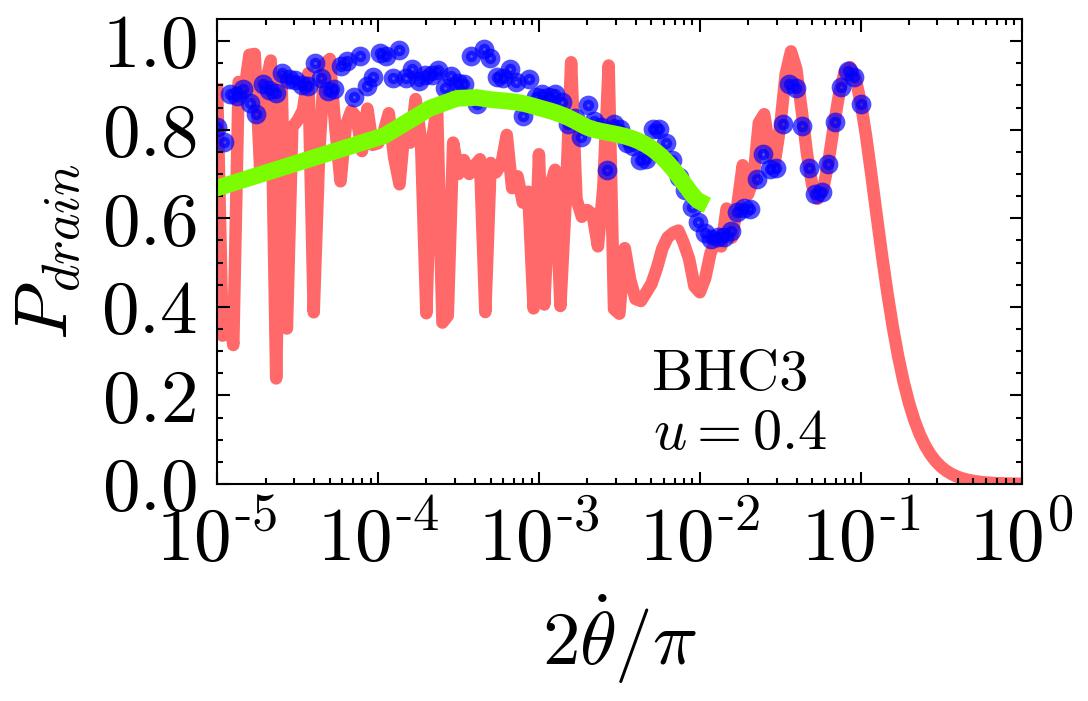}
\includegraphics[width=6cm]{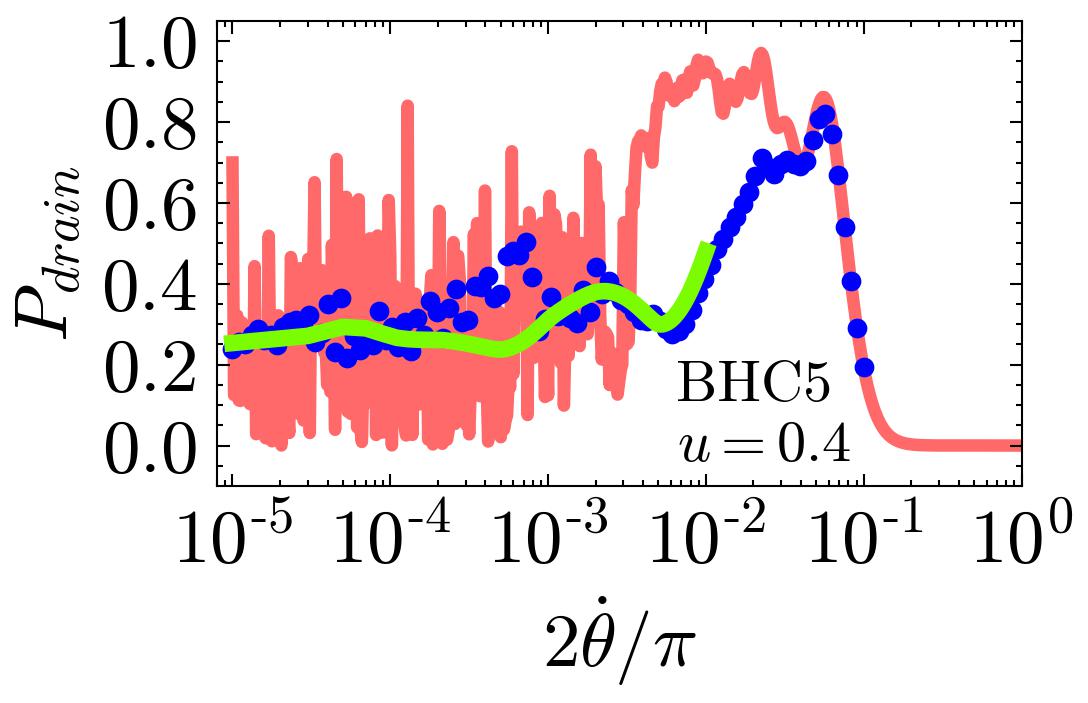}

\caption{\textbf{Transfer efficiency versus sweep rate.} 
First column is for BHC3 with weak and strong interactions: 
${u=0.2, 0.4}$ in the lower and the upper panels respectively. 
Second column is for BHC5 with ${u=0.1,0.4}$.  
The detuning is ${v=0.1}$ in all cases, 
and units of time are chosen such that $K{=}1$. 
Red line represents the MFT result; 
Green line is the \rmrk{{\em averaged}} TWA result; 
and blue points are the QMB results. 
In the first panel, the gray background indicates the diabatic regime.} 
\label{f1}
\end{figure*}

\begin{figure*}
\includegraphics[width=6cm]{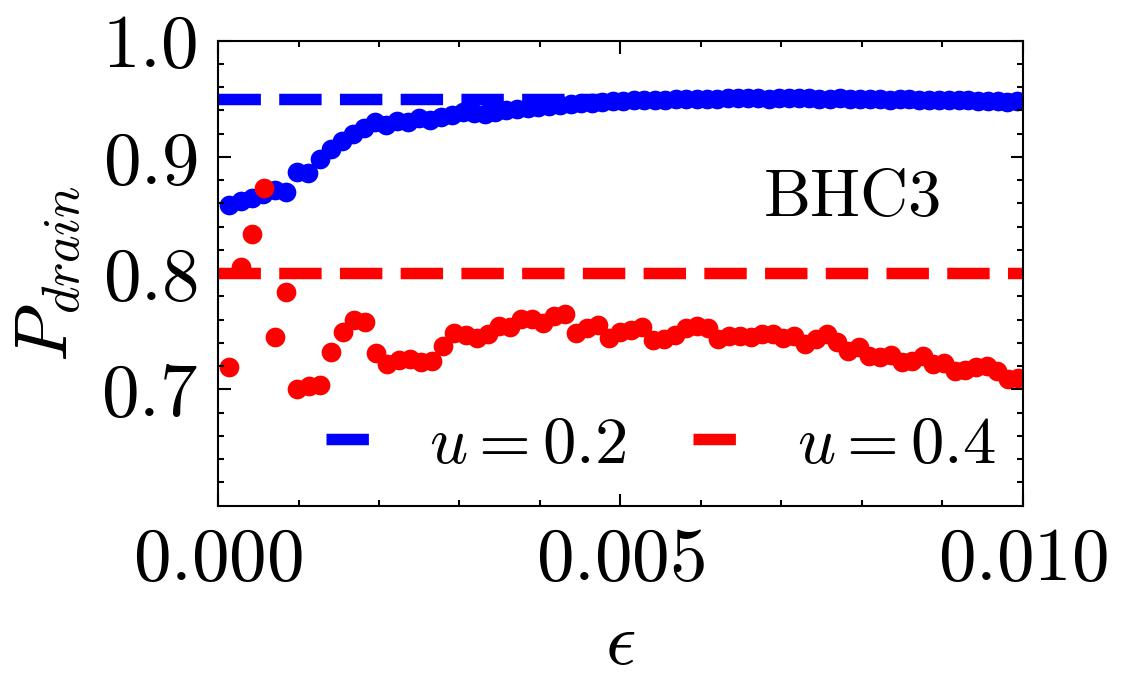}
\includegraphics[width=6cm]{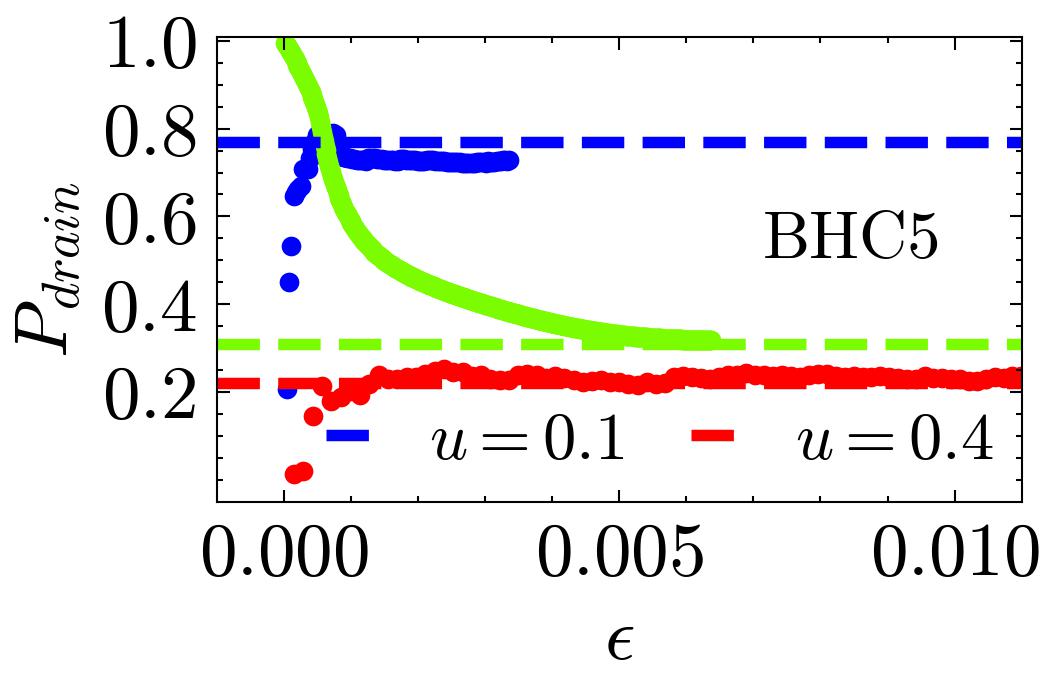}
\caption{\textbf{Breakdown of restricted QCC.} 
The dependence of the average $P_{drain}$ on the energy width $\varepsilon$ of the cloud, whose inverse reflects the number of particles~$N$. It is compared with the quantum expectation value (dashed lines). Left and right panels are for BHC3 and BHC5 with quasistatic sweep ($\dot{\theta} = \pi/2 \times 10^{-5}$). QCC is observed if the width of the cloud is large enough, except the case of BHC3 with ${u=0.4}$. The BHC5 panel includes also a demonstration for the diabatic sweep ($u=0.4$ and $\dot{\theta} = \pi/2 \times 10^{-3}$, green points). 
}
\label{f2}
\end{figure*}

{\bf The BHC model.-}
We consider Bose-Hubbard chains 
\rmrk{with $M$ sites and $N$ particles. This prototype model is well studied theoretically and experimentally \cite{exprBHH1,exprBHH2,atomtronics}.}
We define bose operators $\hat{a}_j$ that annihilate a particle in site~$j$.  The occupations are $\hat{n}_j\equiv\hat{a}_j^\dag\hat{a}_j$, \rmrk{and ${N=\sum_j \hat{n}_j}$ is a constant of motion.} 
The Hamiltonian is  
\begin{eqnarray}
H= \sum_{j=1}^{M}  \left[ 
V_j \hat{n}_j  + \frac{U}{2} \hat{n}_{j}\left(\hat{n}_j{-}1\right)
\right]  \nonumber \\
- \sum_{j=1}^{M-1} \left(\frac{\Omega_{j}}{2}\hat{a}^{\dagger}_{j{+}1} \hat{a}_{j} + H.c.\right)
\rmrk{\, .}
\label{E1}
\end{eqnarray}
The potential is $V_j=V$ at the middle site and zero otherwise. The hopping is ${\Omega_j = K \sin(\theta)}$  at odd~$j$ bonds, and  ${\Omega_j = K \cos(\theta)}$  at even~$j$ bonds. 
The STIRAP-like scheme is realized by launching all particles in site $j{=}1$ and \rmrk{sweeping the control parameter slowly from $\theta{=}0$ to $\theta{=}\pi/2$ with constant rate $\dot{\theta}$}. In the single-particle scenario, the particle is transferred adiabtically from the first site (``source") to the last site (``drain"). In the many body scenario the efficiency is affected by the dimensionless parameters 
\begin{equation}
u = \sqrt{2}\frac{NU}{K}, 
\ \ \ \ \ v= \sqrt{2}\frac{V}{K} 
\rmrk{\, .}
\end{equation} 
\rmrk{Setting non zero ${v\ll 1}$ allows to tune the amount of chaos in the system, but otherwise has no qualitative significance.} The transfer efficiency is defined as 
\rmrk{${P_{drain} = (1/N)\braket{\hat{n}_{j{=}M}}}$} 
at the end of the sweep process.  Numerical results are presented in \Fig{f1} and \Fig{f2} for an $M{=}3$ chain (BHC3) and for an $M{=}5$ chain (BHC5), with interaction $u$ that is either weak or strong. \rmrk{We use the term {\em restricted} QCC to indicate agreement between the {\em averaged} TWA result and the QMB outcome for $P_{drain}$.}

{\bf Adiabatic Sweep process.--} 
\rmrk{In standard single-particle STIRAP with BHC3, see e.g. Section II.B of \cite{stirap_1},  the system follows adiabatically the middle ${E=0}$ orbital 
${ \ket{\alpha} = \cos(\theta)\ket{1}-\sin(\theta)\ket{3}  }$, aka the ``dark state".  In the many body version, in the absence of interactions, $\ket{\alpha}$ is a condensate,  corresponding to a coherent-state in the sense of Perelomov and Gilmore \cite{CSP,CSG}. As explained in the supplementary \cite{SM}, the index $\alpha$ can be regarded as a set of coordinates that indicate a location in phasespace. With non-zero interaction (${u>0}$) the $\alpha$ location of the dark-state is initially (for small $\theta$) a stable elliptic point in the middle of the energy landscape of the BHC3 Hamiltonian. We call it the {\em central stationary point} (SP). In an ideal adiabatic scenario,  the condensate remains localized in the SP, and follows it as $\theta$ is slowly varied.  Had the SP been stable throughout the sweep, the result would be a robust adiabatic population transfer from the first site to the last site. However, for sufficiently large $u$ the SP stability is lost for an intermediate range of $\theta$ values,  hence the transfer efficiency is damaged. }

\begin{figure}
\includegraphics[width=7cm,height=4cm]{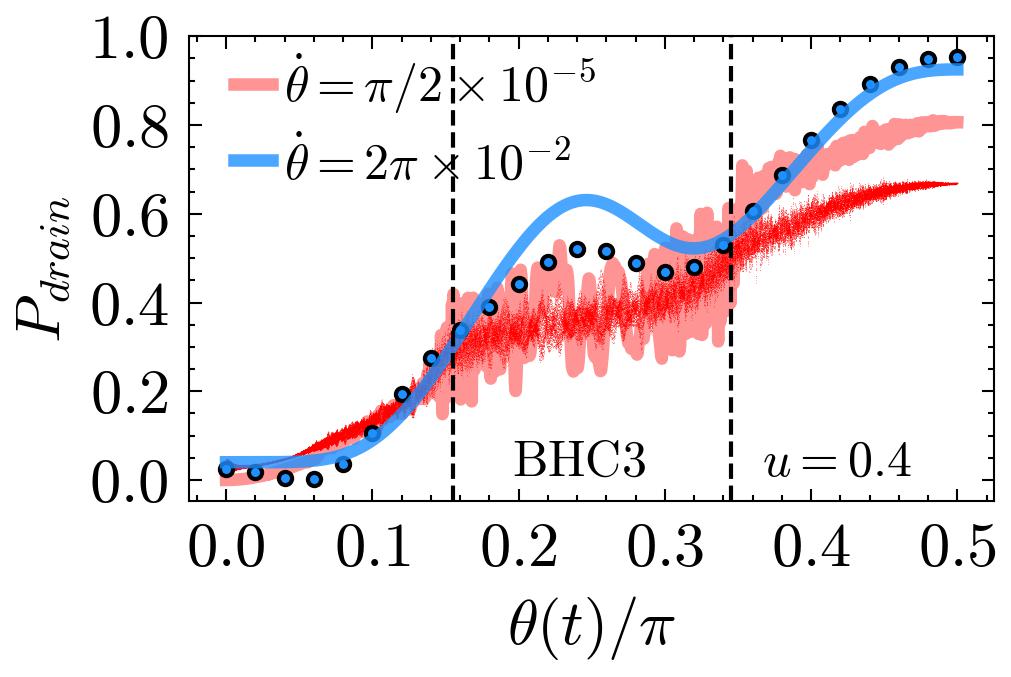}
\caption{\textbf{|Quantum vs Semiclassical Dynamics. } We consider BHC3 with ${u=0.4}$. The evolution of the expectation value $P_{drain}$ is plotted for a cloud of trajectories. In the quasistatic regime (red points) the final average value comes out lower compared with the quantum simulation (faint red line). This is a demonstration of QCC breakdown. In the diabatic regime (blue points and line) the variation is rather smooth, without fluctuations, and restricted QCC is observed. The vertical dashed line indicate the borders of the $\theta$ interval where the SP is unstable.} 
\label{f3}
\end{figure}

\begin{figure}
\centering

\includegraphics[width=8cm,height=4cm]{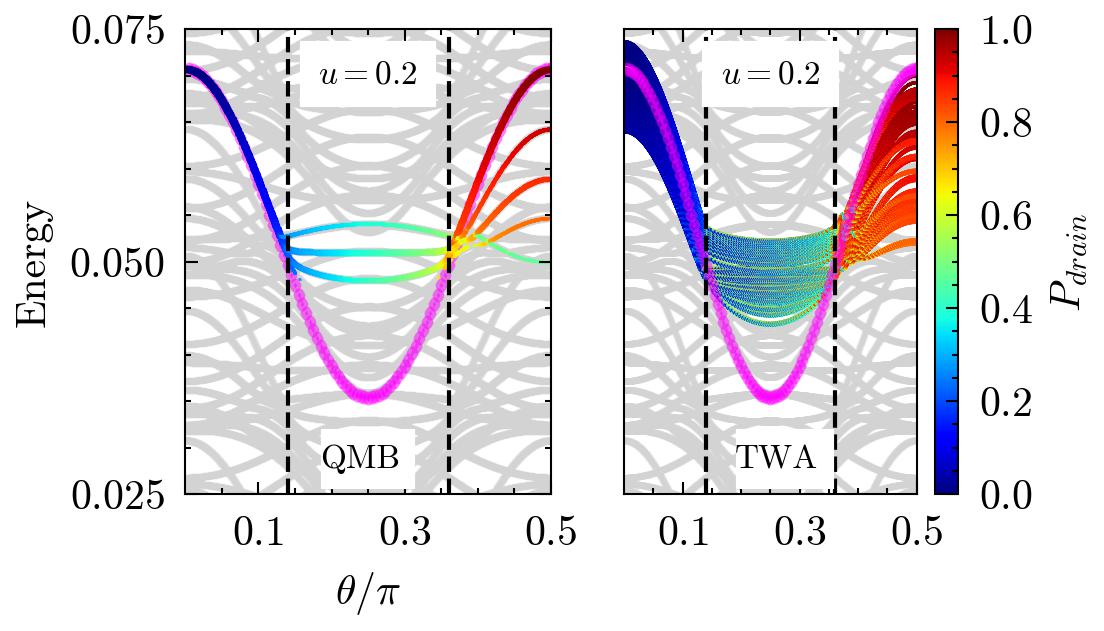}

\includegraphics[width=8cm,height=4cm]{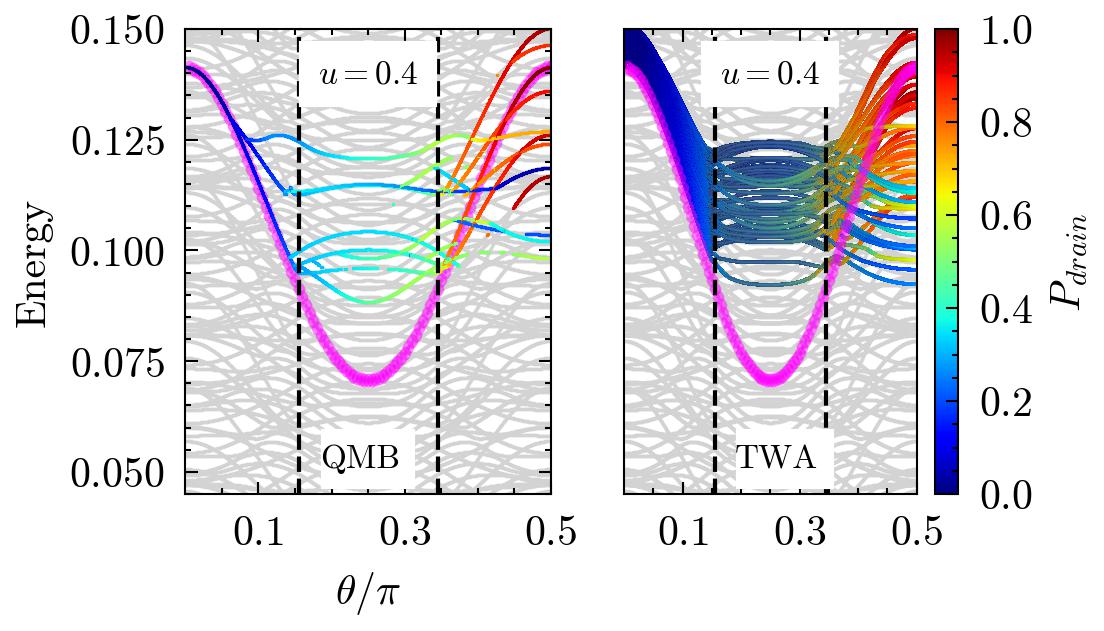}

\caption{ \textbf{Quantum vs semiclassical evolution for BHC3 in the quasistatic regime.} In the background of each panel the adiabatic levels (in gray) are plotted as a function of the $\theta$.  The left panels show the results of quantum simulation. The levels that have non-negligible overlap with the time dependent evolving state are colored. The color-code indicated the parametric evolution of  $P_{drain}$ for each adiabatic level.  The panels on the right display the corresponding classical evolution. It displays the colored points of the corresponding semicalssical cloud.    
The 1st row is for $u=0.2$. 
The 2nd row is for $u=0.4$.
The sweep rate is $\dot{\theta} = \pi/2 \times 10^{-5}$.
Magenta line indicates the energy of the central SP.  
Vertical black dashed lines indicate the borders of the $\theta$ range where the SP is unstable based on Bogoliubov analysis.
}
\label{f4}
\end{figure}

{\bf Sweep rate regimes.--} 
%
%
\rmrk{The stability analysis of the central SP is detailed in the supplementary \cite{SM}, and see \cite{Pethick_Smith_2008,PhysRevE.109.064207}. For any value of $\theta$ we determined the Bogoliubov frequencies $\omega_{k}$ of small-oscillations around the SP. Complex frequency implies that the SP becomes unstable (hyperbolic).  
%
Accordingly, one can distinguish between {\em three} sweep-rate regimes: sudden, diabatic, and quasistatic. 
The sudden-diabatic border is determined by the real part of the Bogoliubov frequencies, namely, 
${ \dot{\theta} \sim \overline{\Re[\omega(\theta)]} }$). The diabatic-quasistatic border is determined by the imaginary part, namely,  
${ \dot{\theta} \sim \overline{\Im[\omega(\theta)]} }$).} 
The optimal transfer efficiency is attained in the diabatic regime (gray area in \Fig{f1}a) where the sweep rate balances between two contradicting demands: on the one hand it should be sufficiently slow with respect to the natural oscillation frequencies, allowing the state of the system to follow the SP; on the other hand it should not be too slow,  or else the system has enough time to spread away from the unstable SP.

{\bf Transfer Efficiency.--} 
The red and blue lines in \Fig{f1} correspond to MFT single classical trajectory and QMB simulations respectively.  The MFT-QMB agreement is obviously poor, notably at the diabatic regime. The MFT predicts almost 100\% efficiency in the diabatic regime, while for finite $N$, which formally corresponds to finite $\hbar$, diabatic efficiency is reduced. We therefore have to adopt a proper TWA semiclassical procedure that takes into account the uncertainty width of the evolving cloud (green line). It turn out that as the cloud's energy variance is increased the transfer efficiency in most cases hits a plateau, becomes independent of the energy width $\varepsilon$, and QCC is established (see \Fig{f2}). The exception is a systematic breakdown of QCC that we witness for BHC3 if $u$ is large: the TWA over-estimates the loss of efficiency. The time-dependent dynamics for this latter case is further demonstrated in \Fig{f3}.  

We conclude that in a realistic experiment, one should expect, in the diabatic regime, a rather sharp crossover from a rather low (green line) to a rather high (red line) efficiency as the number $N$ of trapped particle is increased.  
The other practical observation is that in the extreme quasistatic regime the transfer efficiency can be improved if we are dealing with low dimensional chaos (BHC3), as opposed to longer chains (BHC5) where the slowness does not help. Rather, in the later case, we observe strong $P_{drain}$ fluctuations as $\dot{\theta}$ is decreased.

\begin{figure*}
\includegraphics[width=6.8cm]{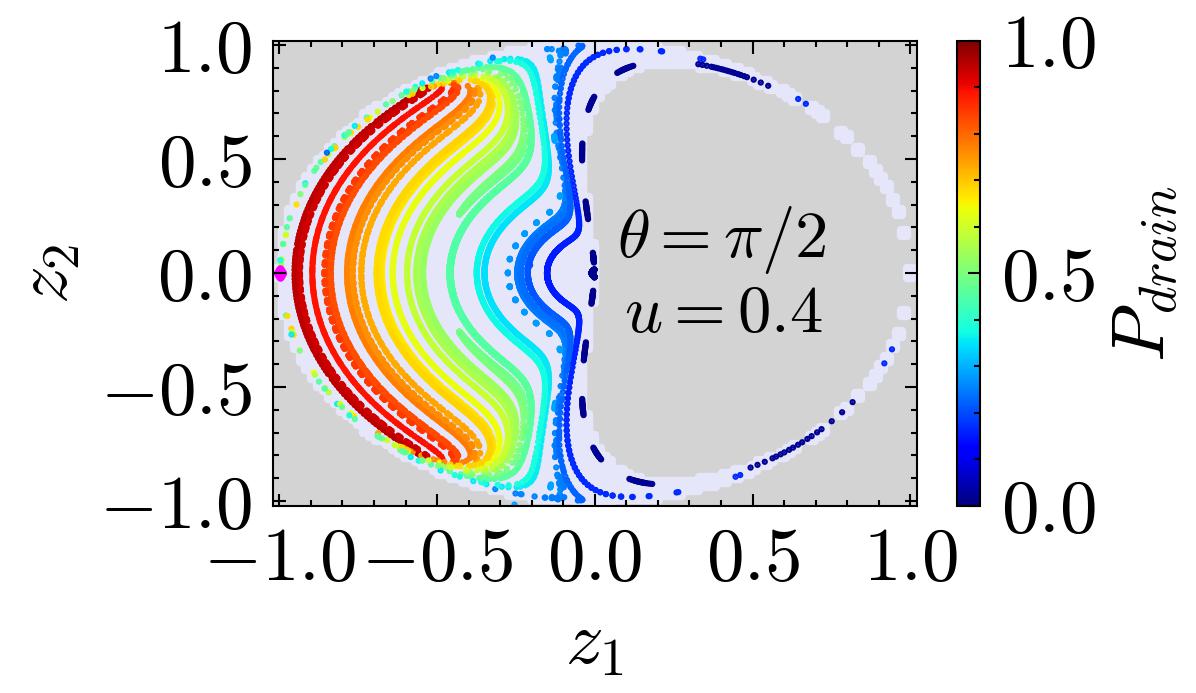}
\includegraphics[width=6.8cm]{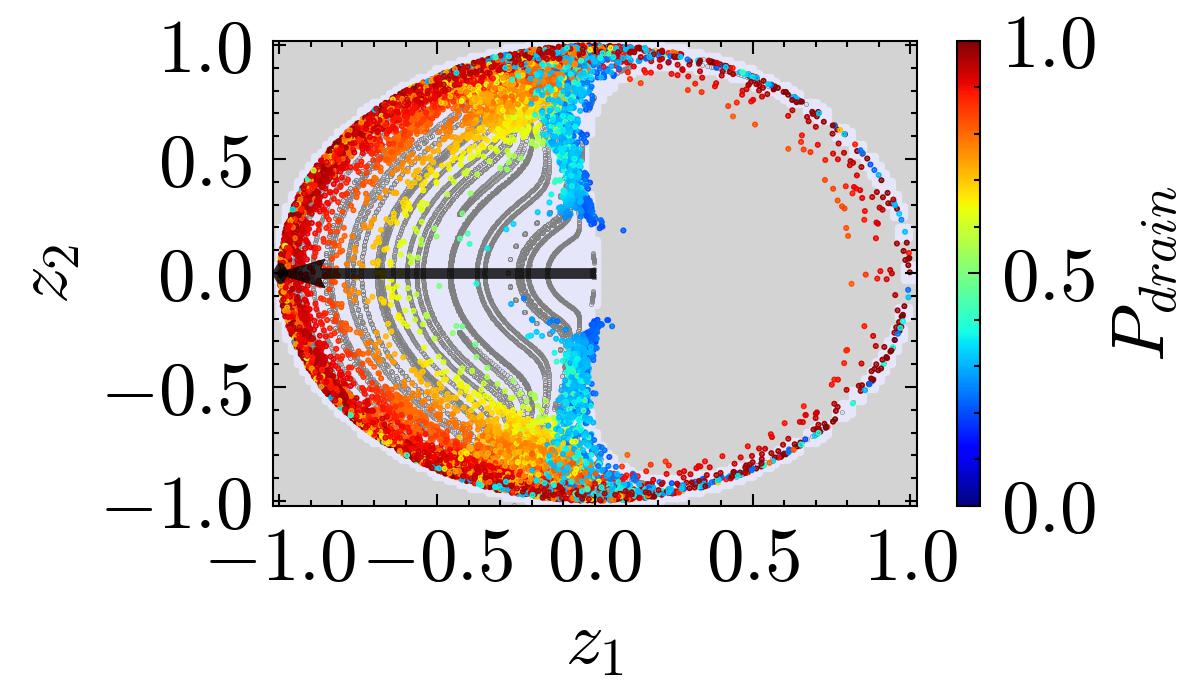} 

\caption{\textbf{Semiclassical inspection of the final state.} 
Left panel is a Poincare section that shows trajectories in phasespace for the $\theta=\pi/2$ Hamiltonian. The color-code is the same as used for the levels in \Fig{f4}. 
The radial coordinate is the first-site depletion $|z|=1{-}(n_1/N)$, 
while the polar coordinate is the conjugate phase variable.
Right panel shows how the points of an evolving cloud are distributed at the end of the ${\dot{\theta}=\pi/2 \times 10^{-5}}$ sweep of 
\cblue{Figs~3~and~4}.
%
Black arrow denotes the motion of the SP to its final position: this motion is followed by the cloud in the diabatic case (not shown). The gray background is a Poincare section of the left panel, which is taken at the SP energy. The points of the cloud do not have exactly the same energy, but still follow the two sets of tori (as implied by their color). Their distribution is not sensitive to the energy. 
}
\label{f5}
\end{figure*}

{\bf Detailed QCC.--} 
\Fig{f3} shows the evolution of the \rmrk{averaged quantity} $P_{drain}$, while \Fig{f4} provides information on the full
probability distribution. One obvious difference between QMB and TWA simulations, is the discretization of the distribution due to energy quantization. \rmrk{Accordingly, when we discuss detailed QCC we focus on the overall envelope of the distribution.} Thus, while the upper panels of \Fig{f4} exhibit excellent detailed QCC, the lower panels demonstrate its breakdown for large $u$. Specifically, in the QMB simulation, unlike the TWA simulation, the energy distribution departs from the energy of the SP {\em before} it becomes unstable. 

High dimensional chaos prevails for any $u$, hence the behavior of the BHC5 system is analogous to that of a BHC3 system that has large $u$.  Indeed we find for BHC5 similar breakdown of detailed QCC (see SM \cite{SM}).  Nevertheless, as expected from the above discussion, we do not witness any systematic breakdown of restricted QCC. Instead we witness strong fluctuations (also wrt $v$, not shown).     

The existence of restricted QCC despite the breakdown of detailed QCC is most conspicuous in the diabatic regime for both BHC3 and BHC5. Our simulations (see SM \cite{SM}) show staggering differences in the detailed energy distribution, yet as we saw in \Fig{f1} the quantum and semiclassical transfer efficiencies in the diabatic regime are the same.

{\bf Spreading mechanism.--} 
In order to explain our numerical findings we use a semiclassical picture. The explanation below is based on an extensive numerical exploration that is documented in the SM \cite{SM}.  
Consider a quasistatic sweep with BHC3. The followed SP is initially a stable island in phasespace. Once the SP become unstable, the cloud can spread along the newly formed torus that contains the SP. We denote this intra-torus spreading as {\em longitudinal} spreading. After the onset of instability, the cloud follows adiabatically this new torus, that departs from the SP. Ideally, in the quasistatic limit, this torus collapse back into the SP in the \rmrk{second-half} of the sweep process. This may be viewed as a classical shortcut to adiabaticity, where the system shuttles on the new torus while the SP is unstable.

The semiclassical cloud can also expand into other tori due to the non-linearity~$u$. This {\em transverse} spreading is dominant for large~$u$, and damages adiabaticity. Furthermore, due to the formation of a stochastic strip the quasistatic limit is no longer adiabatic. At the end of the sweep the cloud is distributed over two sets of tori, see \Fig{f5}. This distribution is not sensitive to the energy, which explains the findings in \Fig{f2}.

Quantum mechanically, partial longitudinal spreading corresponds to superposition with far away energy levels, i.e.  to jump-like  transitions (see \Fig{f2b} of the SM), conforming to the Fermi golden rule (FGR) paradigm. By contrast, transverse transitions are  predominantly of the Landau-Zener (LZ) type \cite{LZ_1,LZ_2,Nikitin1984}. They appear at avoided crossings (no jumps) and do not obey QCC. Furthermore, they can occur before the stability of the SP is lost, which can be regarded as dynamical tunneling.  

In the diabatic regime detailed QCC is lost due to the longitudinal-spreading-induced jumps (see SM \cite{SM}). However, restricted QCC is maintained due to the perturbative nature of the FGR-like transitions, 
as explained in \cite{HILLER20061025} and reference therein.  
In the high dimensional chaos of BHC5 the connected topology of phasespace implies that the distinction between longitudinal and transverse spreading is blurred. Consequently the weight of LZ-transitions become smaller, and restricted QCC is restored.

{\bf Summary.--} 
The design of a manybody transfer protocol has to take into account the fingerprints of non-linearity and chaos that dominates the underlying phasespace dynamics. In the diabatic regime, there is a rather sharp transition from low to high efficiency as the number of Bosons exceeds a threshold value. TWA, unlike MFT, can be trusted in general, but is quantitatively broken in the quasistatic regime for low dimensional chaos. In general, adiabaticity is spoiled in the quasistatic limit, but is recovered in the case of weak low-dimensional nonlinearity.


{\bf Acknowledgment.-- } 
DC acknowledges support by the Israel Science Foundation (Grant No. 518/22).
AV acknowledges support from the NSF through a grant for ITAMP at Harvard University.



%

\clearpage
\onecolumngrid
\pagestyle{empty}

\renewcommand{\thefigure}{S\arabic{figure}}
\setcounter{figure}{0}
\renewcommand{\theequation}{S-\arabic{equation}}
\setcounter{equation}{0}

\begin{center}
{\large \bf Many-body adiabatic passage: \\
instability, chaos, and quantum classical correspondence}

\ \\

{Anant Vijay Varma, Amichay Vardi, and Doron Cohen}

\ \\

{\large (Supplementary Material)}

\end{center}

In this Supplementary we provide details on the SP stability analysis; including how chaos is identified; and the determination of the quasistatic-diabtic border based on the Bogoliubov analysis. We also provide additional panels that illustrate the quantum and the semiclassical evolution; and in particular further clarify the spreading-dynamics of the cloud in phasespace.  

\ \\ 

\sectA{Phasespace picture for bosons}

The harmonic oscillator is described by canonical coordinates $(q,p)$, and one defines destruction operator 
$\hat{a}_0 = 2^{-1/2} (q+ip)$. The coherent state 
$\alpha = 2^{-1/2} (\bar{q}+i\bar{p})$ 
is defined as the phasespace translation $D(\alpha)$ of the ground state $\ket{0}$, namely $\ket{\alpha} = D(\alpha) \ket{0}$. One can define a destruction operator 
${\hat{a}_{\alpha} = \hat{a}_0 - \alpha}$ around the $\alpha$ phasespace location. In the Heisenberg picture  
${D^{\dag} \hat{a}_0 D = \hat{a}_0+\alpha }$, which can equivalently written as ${D \hat{a}_0 D^{\dag} = \hat{a}_{\alpha} }$.
Considering general time evolution that is generated by some Hamiltonian, the leading semiclassical approximation is analogously written as  
\beq \label{eU}
U(t) \, \hat{a}_{\alpha(0)}^{\dag} \, U(t)^{\dag} \ \ \approx \ \ \hat{a}_{\alpha(t)}^{\dag}
\eeq
where $\alpha(t)$ satisfies Hamilton's equations of motion. In view of the manybody generalization below we referred here to the creation operator $a^{\dag}$.

Perelomov and Gilmore \cite{CSP,CSG} have generalized the concept of coherent states in a way that allows application in the manybody context. Considering a system of bosons with creation operators $\hat{a}_j^{\dagger}$, we define an operator that creates particle in orbital $\alpha$, namely, 
\beq
\hat{a}_{\alpha}^{\dagger} \ = \ \sum_{j=1}^M \alpha_j \hat{a}_j^{\dagger}, 
\ \ \ \ \ \text{where} \ \sum_{j=1}^M |\alpha_j|^2 =1
\eeq
Then we can generate a set of coherent states as follows: 
\beq
\ket{\alpha} \ = \ \frac{1}{\sqrt{N}} \left[\hat{a}_{\alpha}^{\dagger} \right]^N \ket{\text{vacuum}}  
\eeq
All those $\alpha$ states can be regarded as SU($M$) ``rotations" of an arbitrarily selected coherent state $\alpha_0$. This is completely analogous to the harmonic oscillator, where all the $\alpha$ states are ``translations" of each other.  
Furthermore, if $U(t)$ operates on a given coherent state preparation $\alpha(0)$, then, using \Eq{eU}, we deduce the semiclassical approximation 
\beq
U(t) \ket{\alpha(0)} \ \approx \ \ket{\alpha(t)} 
\eeq
where $\alpha(t)$ satisfies Hamilton's equations of motion. This is what we called MFT: the manybody evolution is derived from a single classical trajectory that is generated by the classical Hamiltonian. 

We use the following terminolgy: $\alpha$ indicates phase space location; and $\ket{\alpha}$ is a coherent state that is supported by $\alpha$. Such state describes condensation of particles in a single {\em orbital} (single particle state) that is parametrized by $\alpha$. For a two site system (dimer) the common paramterization is 
\beq
\alpha_{\text{dimer}} \ = \ \left( \cos(\theta)e^{-i\varphi/2}, \sin(\theta)e^{i\varphi/2} \right),
\eeq
hence phasespace is the Bloch sphere, with spherical coordinates $(\theta,\varphi)$ that describe the population imbalance and the relative phase. For a trimer we need two pairs of conjugate coordinates. Of particular interest is the so-called ``dark state" whose phasespace location is 
\beq
\alpha_{\text{trimer}} \ = \ \left( \cos(\theta), 0, -\sin(\theta) \right) 
\ \ \equiv \ \ \alpha_{\text{SP}}
\eeq
This is the zero-energy eigenstate of the single particle BHC3 Hamiltonian. In the single particle STIRAP scenario the system follows adiabatically this orbital -- see e.g. Section II.B of \cite{stirap_1}. In the many body context, in the absence of inter-particle interactions, it is the coherent state $\ket{\alpha_{\text{SP}}}$ that is being ``rotated".  From a semiclassical perspective $\alpha_{\text{SP}}$ indicates an SP of the dynamics that is being followed. We call it the central SP, because it is located at the middle of the spectrum. In the next section we provide stability analysis of the SP.

\sectA{The central SP}

In the classical limit, the field operators $\hat{a}_{j}$ can be replaced by complex numbers  $\alpha_{j} = \sqrt{n_{j}} e^{i \varphi_{j} }$,
with ${j=1,\cdots,M}$. Thus, the classical motion has $M$ degrees of freedom, with $\{{n_{j},\varphi_{j}}\}$  serving as conjugate action-angle variables. Owing to the U(1) symmetry, the classical phase space can be further reduced to ${d=M{-}1}$ degrees of freedom. 
For demonstration let us consider BHC3. A possible choice of canonical variables is $p_{1} = n_{1}/N$, $p_{2} = n_{2}/N$, $q_{1} = \varphi_{1}-\varphi_{3}$, and $q_{2}=\varphi_{2}-\varphi_{3}$, resulting in the classical Hamiltonian:
\begin{eqnarray}
\dfrac{H}{N} = && V \ p_{2}+  \frac{NU}{2} \ (p_{1}^{2} + p_{2}^{2} + (1{-}p_{1}{-}p_{2})^{2} )
\label{E6} \\ \nn
&& -  \Omega \left( \sqrt{p_1 p_2} 
\cos (q_{1}{-}q_{2}) +\sqrt{p_{2} (1{-}p_{1}{-}p_{2})} \cos q_{2} \right)
\end{eqnarray}
This classical Hamiltonian has a mid-spectrum fixed-point whose coordinates are $p_1{=}1/2$ , $q_1{=}\pi$, $p_2{=}0$ while $q_2$ is ill defined. In terms of the original canonical variables its location is 
${\alpha}_{\text{SP}}=(1/\sqrt{2},0,-1/\sqrt{2})$. 
Its energy is
\begin{equation} \label{eSP}
E_{SP} = \dfrac{1}{4}N^2U 
\end{equation}
Quantum mechanically, this SP supports a coherent state where all particles occupy the dark-state orbital, namely,   
\begin{equation}
|\alpha_\text{SP}\rangle = \dfrac{1}{\sqrt{2^{N} N!}} (\hat{a}^{\dagger}_{1} -\hat{a}^{\dagger}_{3})^{N} \ket{0}~.
\end{equation}
The mid-spectrum SP remains a stationary point of the classical dynamics even in the presence of interaction. It is a fixed point of the discrete nonlinear Schr\"{o}dinger equation \cite{SM_1,Pethick_Smith_2008}.  By contrast, the SP-supported coherent state is an exact eigenstate of the many-body Hamiltonian only for $U=0$.  

\sectA{SP stability analysis}

Stability analysis means that we find the frequencies $\omega_k$ of the small oscillations around the SP.  These are known as the Bogoliubov frequencies. If all these frequencies are real and positive it means that the SP is a stable minimum. If some are negative (but real) it indicates that the SP is a stable elliptic point that seats on a saddle in the energy landscape.  If some of the frequencies become complex, it indicates that the SP is an unstable hyperbolic point, and possibly the emergence of chaos in its vicinity. 

We clarify the stability analysis of the central SP for BHC3. We follow the presentation of \cite{PhysRevE.109.064207}. The generalization to to BHC5 is straightforward and not presented explicitly for obvious pedagogical reason. But we provide the bottom line results, for both BHC3 and BHC5, and for a wide range of model parameters.   

The classical SPs are found by solving:
\begin{equation}
i \dot{\bm{\alpha}} = (H_{0} + u \mathcal{P}) \bm{\alpha} = \mu \bm{\alpha},
\label{A1}  
\end{equation}
Above we adopt units of time such that $\Omega=1$. 
The matrices $H_{0}$ and $\mathcal{P}$ are:
\begin{equation}
H_{0} = \begin{pmatrix}
 0 & -\frac{\sin\theta}{2}  & 0 \\ 
-\frac{\sin\theta}{2} & v &  -\frac{\cos\theta}{2} \\ 
0  & -\frac{\cos\theta}{2} &  0
\end{pmatrix}, 
\ \ \ \ \ \ 
\mathcal{P} = \begin{pmatrix}
p_{1} & 0 & 0\\ 
0 & p_{2} & 0\\ 
0 & 0 & p_{3}
\end{pmatrix},
\label{A2}   
\end{equation}
where $p_{j}= |\alpha_{j}|^{2}$. The dynamical stability analysis of this SP is carried out via diagonalization of the Bogoliubov matrix:
\begin{equation}
\begin{pmatrix}
H_{0} + 2 u \mathcal{P}- \mu & -u \mathcal{P}   \\ 
u \mathcal{P}  & -(H_{0} + 2 u \mathcal{P}- \mu)  
\end{pmatrix}
\label{A3}
\end{equation}
resulting in 3 pairs of characteristic frequencies $\pm\omega_{k}$. 
We provide explicit expressions for $\theta=\pi/4$. 
The dark-state SP is $\alpha_{\text{SP}}=(1/\sqrt{2},0,-1/\sqrt{2})$. 
The Bogoliubov frequencies are indexed by ${k=\{0,+,-\}}$. The trivial frequency $\omega_{0}=0$ is implied by conservation of particles, while
\beq 
&& \textstyle \nonumber
\omega_{+} = \frac{\sqrt{\sqrt{\left((u-2 v )^2+4\right)^2-16 \left(u^2-2 u v +1\right)}+u^2-4 u v +4 v ^2+4}}{2 \sqrt{2}}  
\\ && \textstyle \nonumber
\omega_{-} = - \frac{\sqrt{-\sqrt{(u-2 v ) \left(u^3-6 u^2 v +4 u \left(3 v ^2-2\right)-8 v  \left(v ^2+2\right)\right)}+(u-2 v )^2+4}}{2 \sqrt{2}} 
\eeq
For clarity we restore the units of $\Omega$, 
and write the expression for $v=0$, 
\beq
\omega_{\pm} \ \ = \ \ \pm \frac{\Omega}{2\sqrt{2}} 
\left[ (4+u^2) \pm u\sqrt{u^2-8} \right]^{1/2}
\label{A4}
\eeq
Setting $u=0$ one can easily verify that the Bogoliubov frequencies ${\omega_{\pm} = \pm \Omega/\sqrt{2}}$ for the non-interacting system correspond to single-particle  transitions to the upper or to the lower orbitals.

\begin{figure}[tbh!]
\includegraphics[width=6cm]{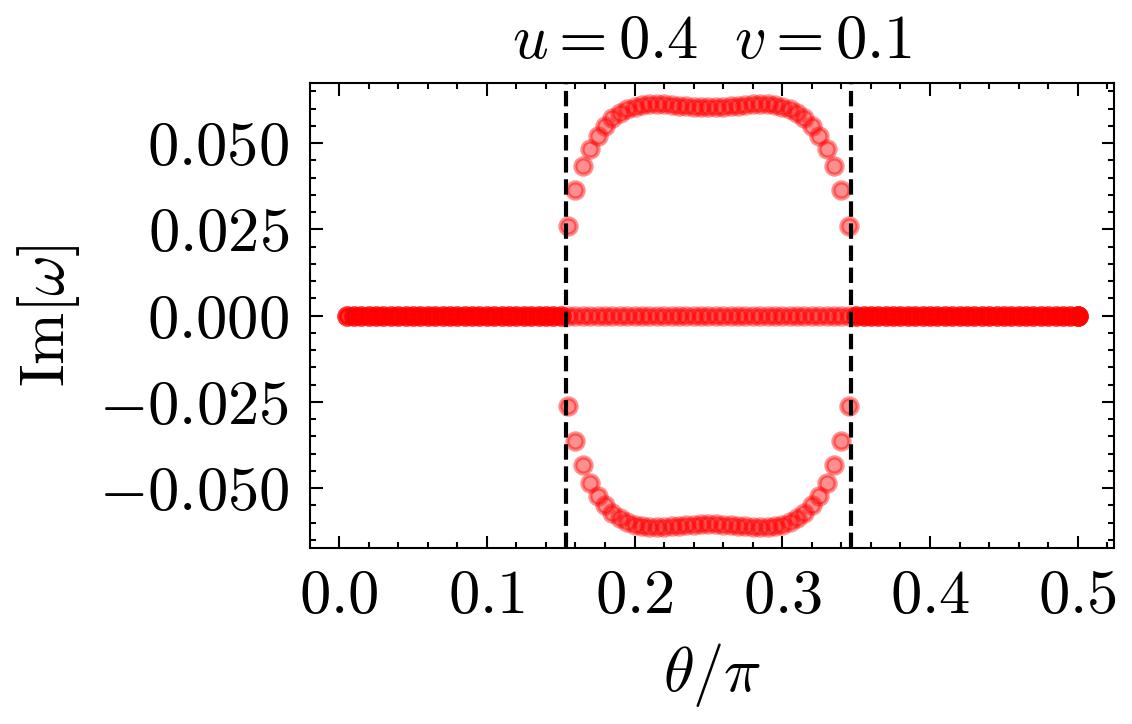}
\includegraphics[width=6cm]{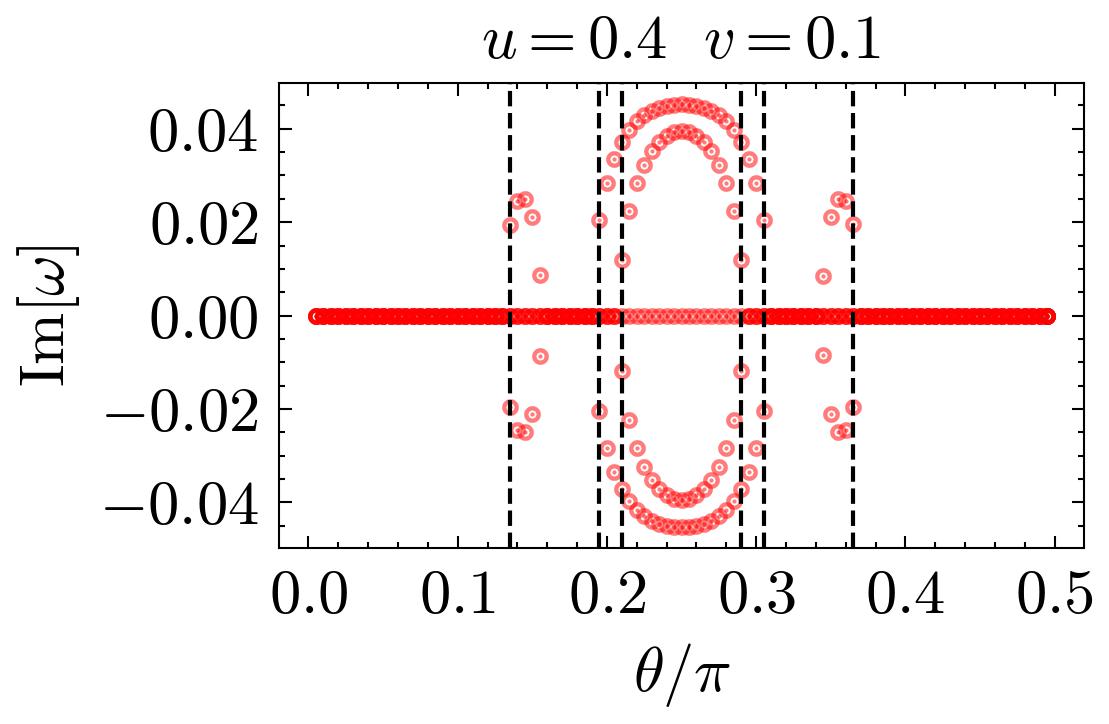}
\caption{\textbf{Bogoliubov Analysis:} 
The result for $\text{Im}(\omega)$ of the SP as a function of $\theta$. Left and right panels are for BHC3 and BHC5 respectively.}
\label{fig3_2}
\end{figure}

\begin{figure}[tbh!]
\includegraphics[width=7cm]{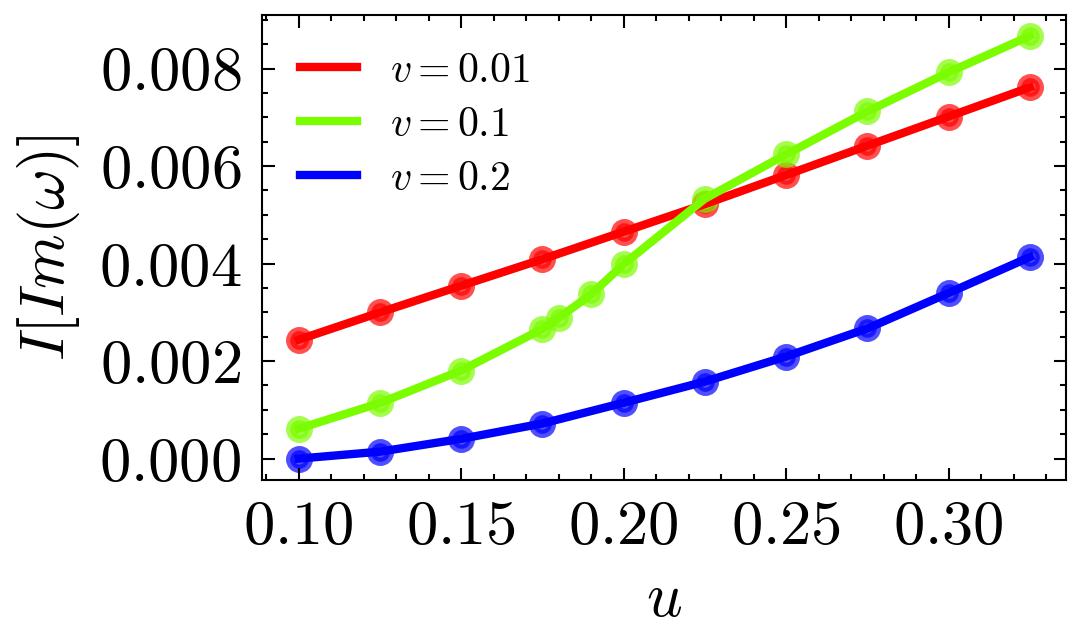}
\caption{\textbf{BHC3:} Time-integrated Im[$\omega$] versus $u$.}
\label{fig5_2}
\end{figure}

\begin{figure}[tbh!]
\centering
\includegraphics[width=0.45 \columnwidth]{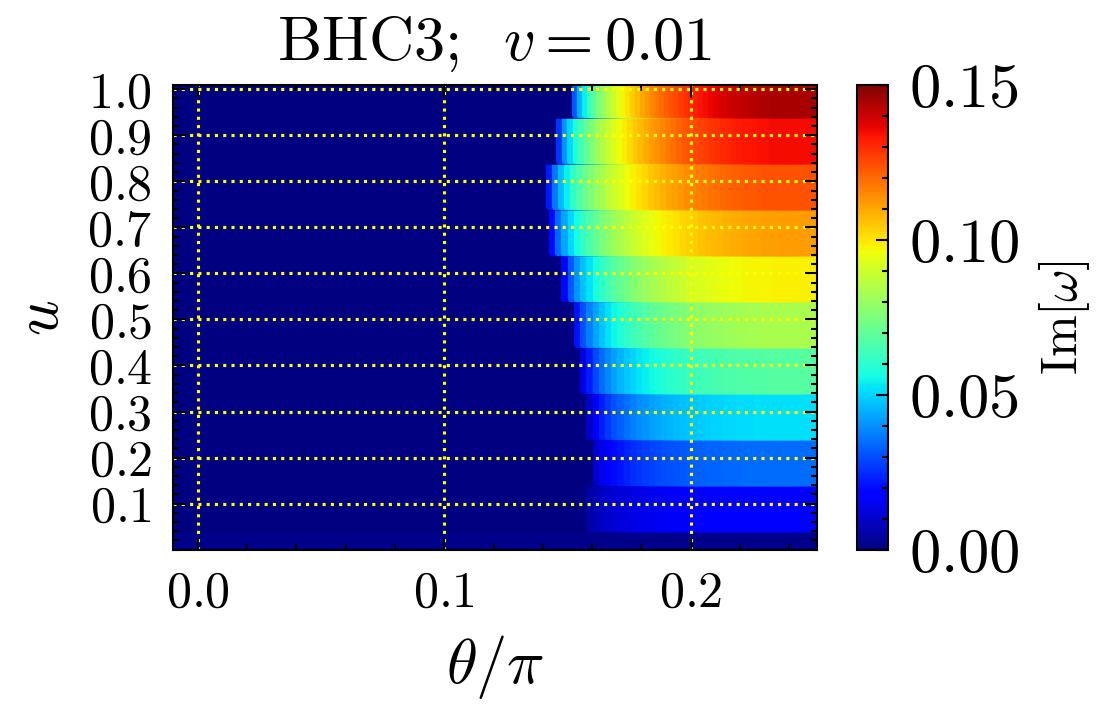}
\includegraphics[width=0.45 \columnwidth]{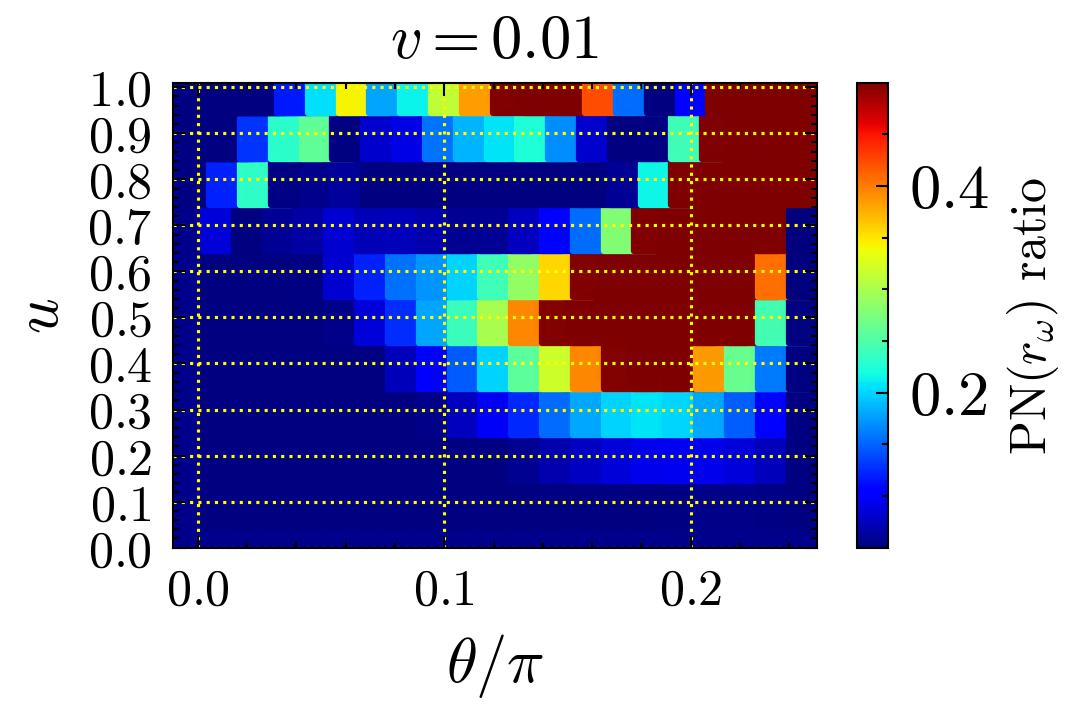}
\includegraphics[width=0.45 \columnwidth]{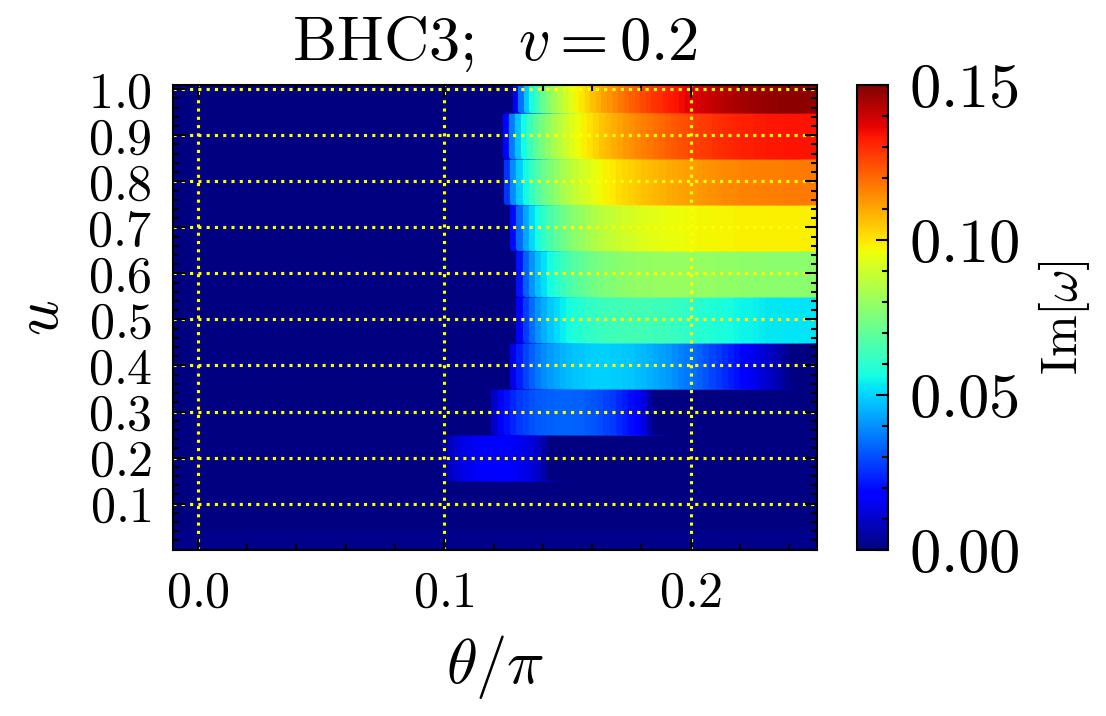}
\includegraphics[width=0.45 \columnwidth]{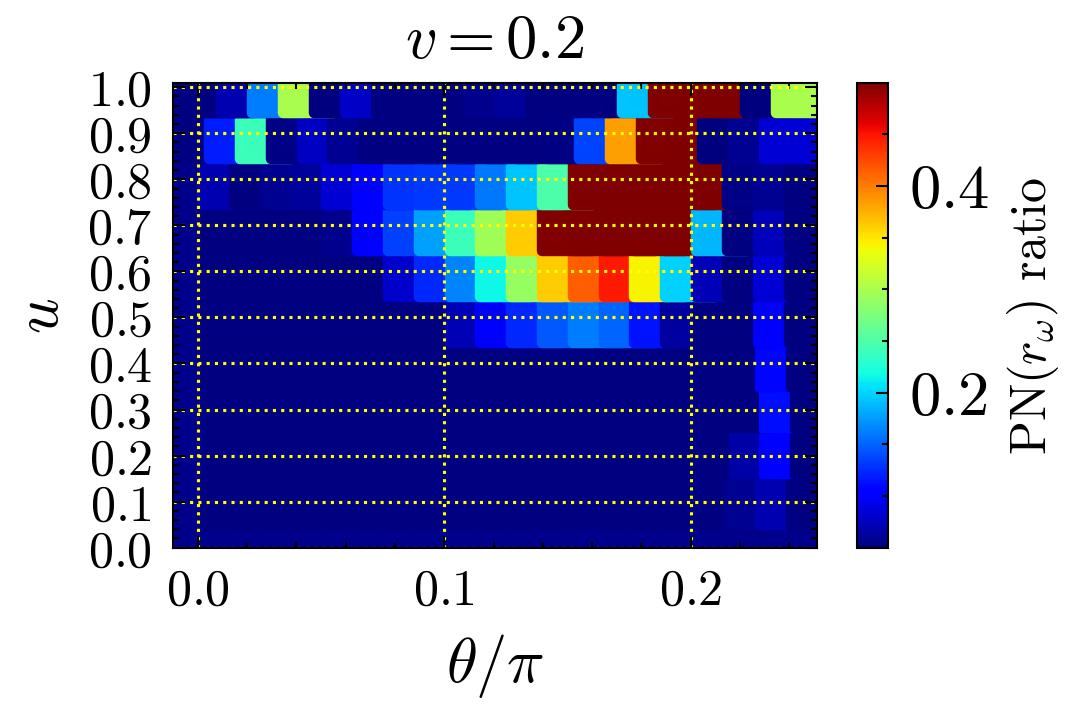}
\includegraphics[width=0.45 \columnwidth]{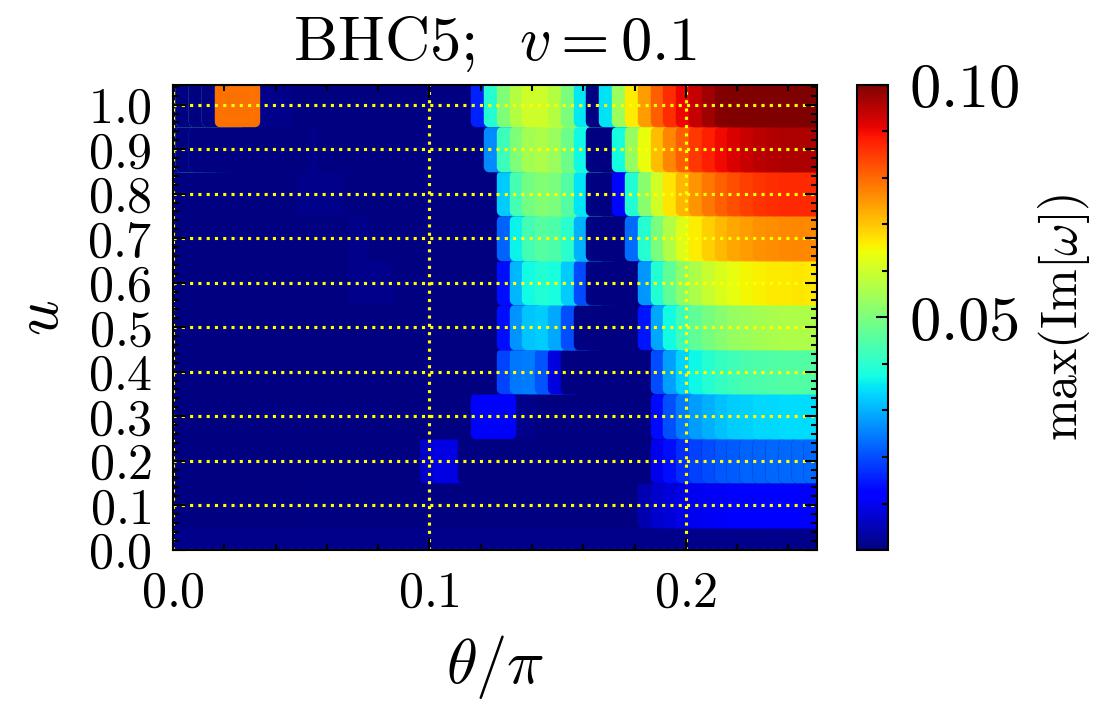}
\includegraphics[width=0.45 \columnwidth]{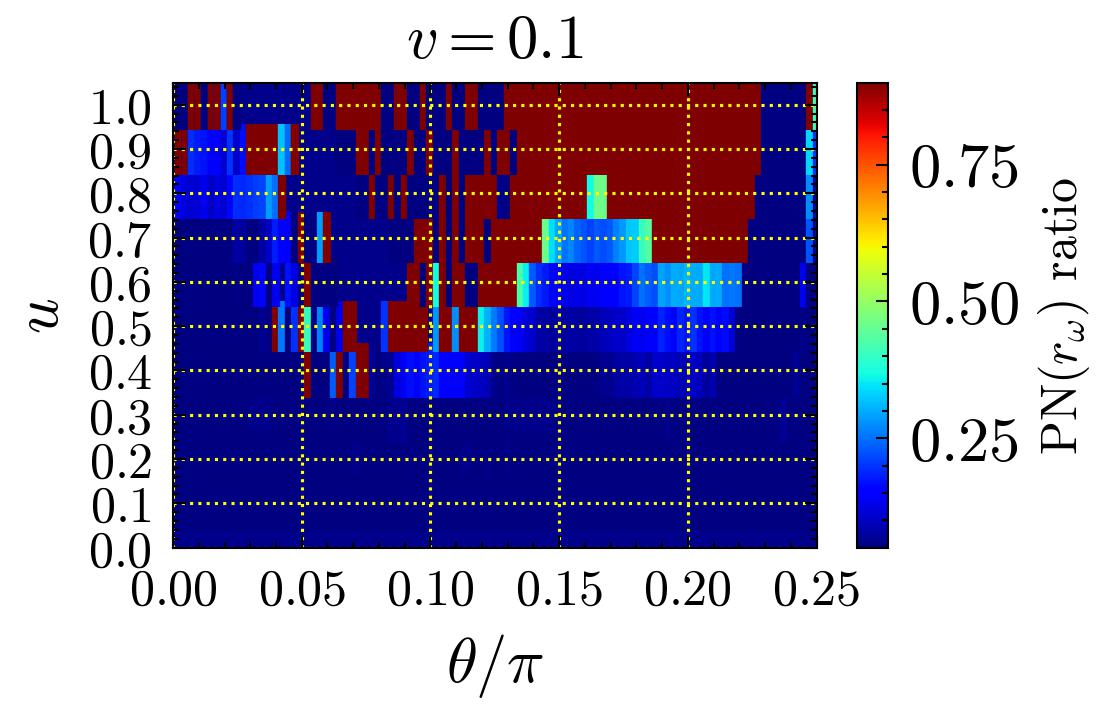}
\caption{\textbf{Bogoliubov and chaos analysis.} 
Left: Bogoliubov analysis. Right: chaos analysis. 
Misc $v$ values are used for BHC3 and BHC5, 
which provides some control over the instability.
Chaos analysis is based on Fourier transform of $r(t)$, the distance from the SP. 
} 
\label{fig3_1}
\end{figure}

In \Fig{fig3_2} we plot $\text{Im}(\omega_k)$ versus $\theta$ for both BHC3 and BHC5. Such plots allow to determine the range where the SP becomes unstable. The time-integrated $\text{Im}(\omega)$, denoted as $I[\text{Im}(\omega)]$, is plotted in \Fig{fig5_2} versus $u$. This can serve as a measure for the departure of the cloud from the SP during the sweep process. 
In the left panels of \Fig{fig3_1}, each row of the image provides  $\text{Im}(\omega)$ for a different value of~$u$. This is compared (right panels) with what we call ``chaos analysis".  The latter is based on characterization of  trajectories $r(t)$ that are launched at the vicinity of the origin ($r=0$) at the beginning of the sweep. Here $r$ is the distance from the SP. The chaos measure is the participation number $\text{PN}(r_{\omega})$ of the power spectrum $|r_{\omega}|^2$ that is obtained via Fourier transform of $r(t)$. 

Misc values of the detuning $v$ are used for BHC3 and BHC5, 
which provides some control over the instability. Note that due to the control over the detuning we can shift the minimum value of $u$ which allows instability and chaos. The small value of $u$ for BHC3 has been chosen such that we have instability while chaos is negligible.

\clearpage
\sectA{The quasistatic-diabtic border}

\begin{figure}[tbh!]
\includegraphics[width=5cm]{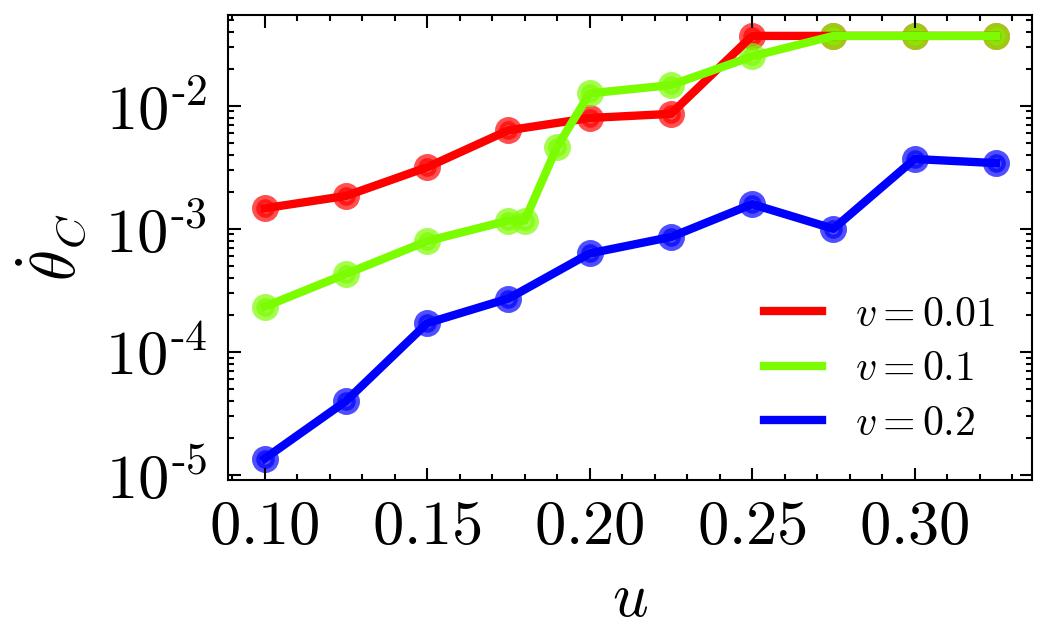}
\includegraphics[width=5cm]{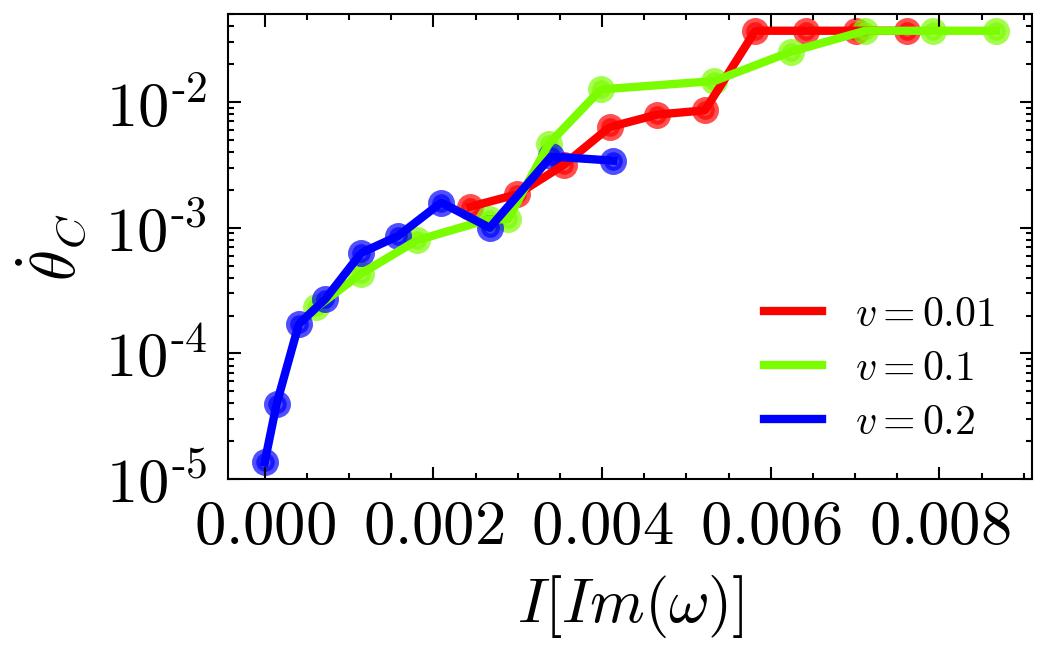}

\includegraphics[width=5cm]{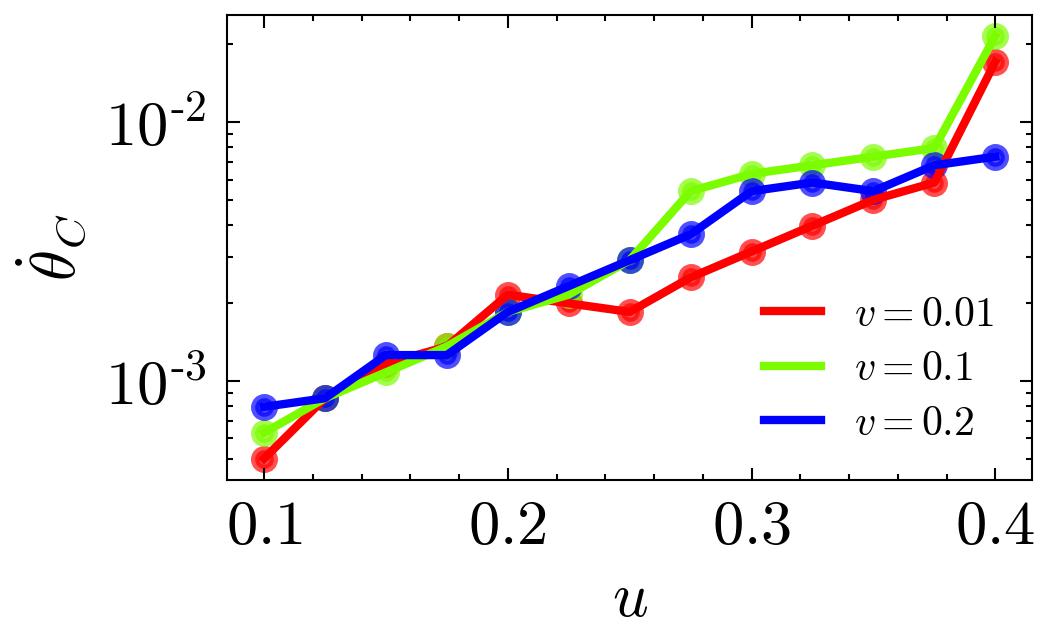}
\includegraphics[width=5cm]{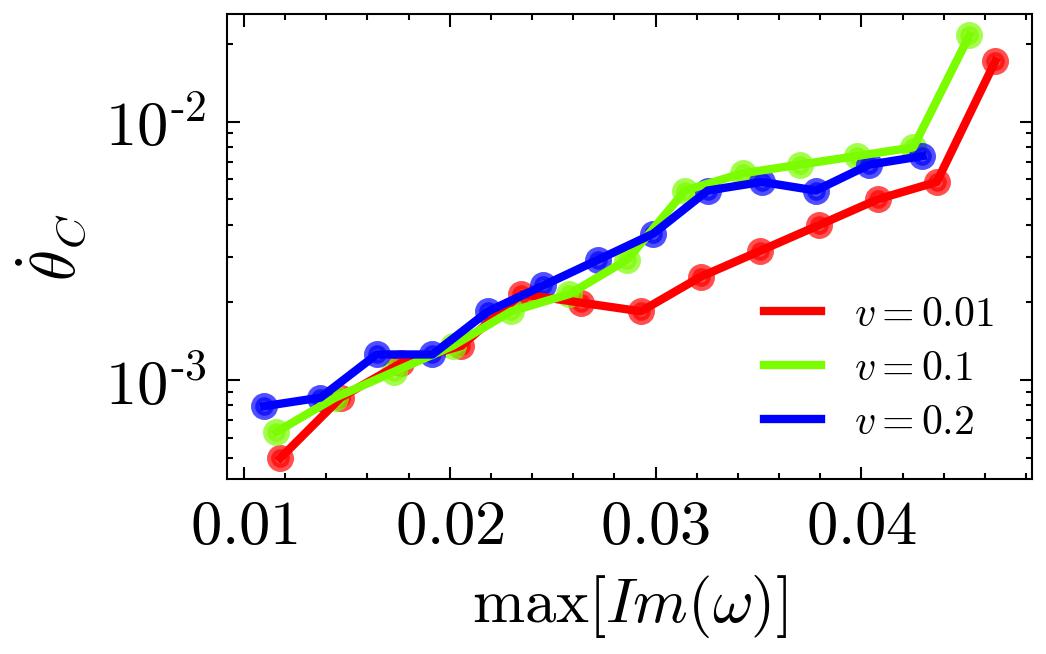}

\caption{\textbf{The quasistatic-diabtic border.} 
In the left panels the border $\theta_c$ is plotted versus $u$ for BHC3 (upper panel) and for BHC5 (lower panel). In the associated right panels this border is plotted versus $I[\text{Im}(\omega)]$ and versus $\max[\text{Im}(\omega)]$ respectively.  
}
\label{fig5_3}
\end{figure}

There are 3 regime of sweep rate: quasistatic; diabatic; sudden. In the diabatic regime the cloud does not have the time to spread in the unstable direction, and manages to follows the SP. Consequently, the transfer efficiency is close to unity. The quasistatic-diabatic border $\dot{\theta}_c$ is determined by the imaginary part of the Bogoliubov frequency.  To be more precise it is determined by the time-integral $I[\text{Im}(\omega)]$, as demonstrated for BHC3 in \Fig{fig5_3}. For BHC5 the stability analysis is multi-dimensional and fragmented, and therefore instead of plotting against the (ill defined) integral we plot versus $\max[\text{Im}(\omega)]$.     

\ \\

\sectA{Chaos analysis for the driven BHC}

The identification of chaos vs $\theta$ is based on simulations with frozen Hamiltonians.  In the time dependent problem we want to verify that the overall dynamics that is generated by the time-dependent Hamiltonian is chaotic in some sense. The energy is no longer constant of motion, so we plot for a cloud of trajectories the final energy versus the initial energy and verify that they are not correlated.   

\ \\

\begin{figure}[tbh!]
\includegraphics[width=5cm]{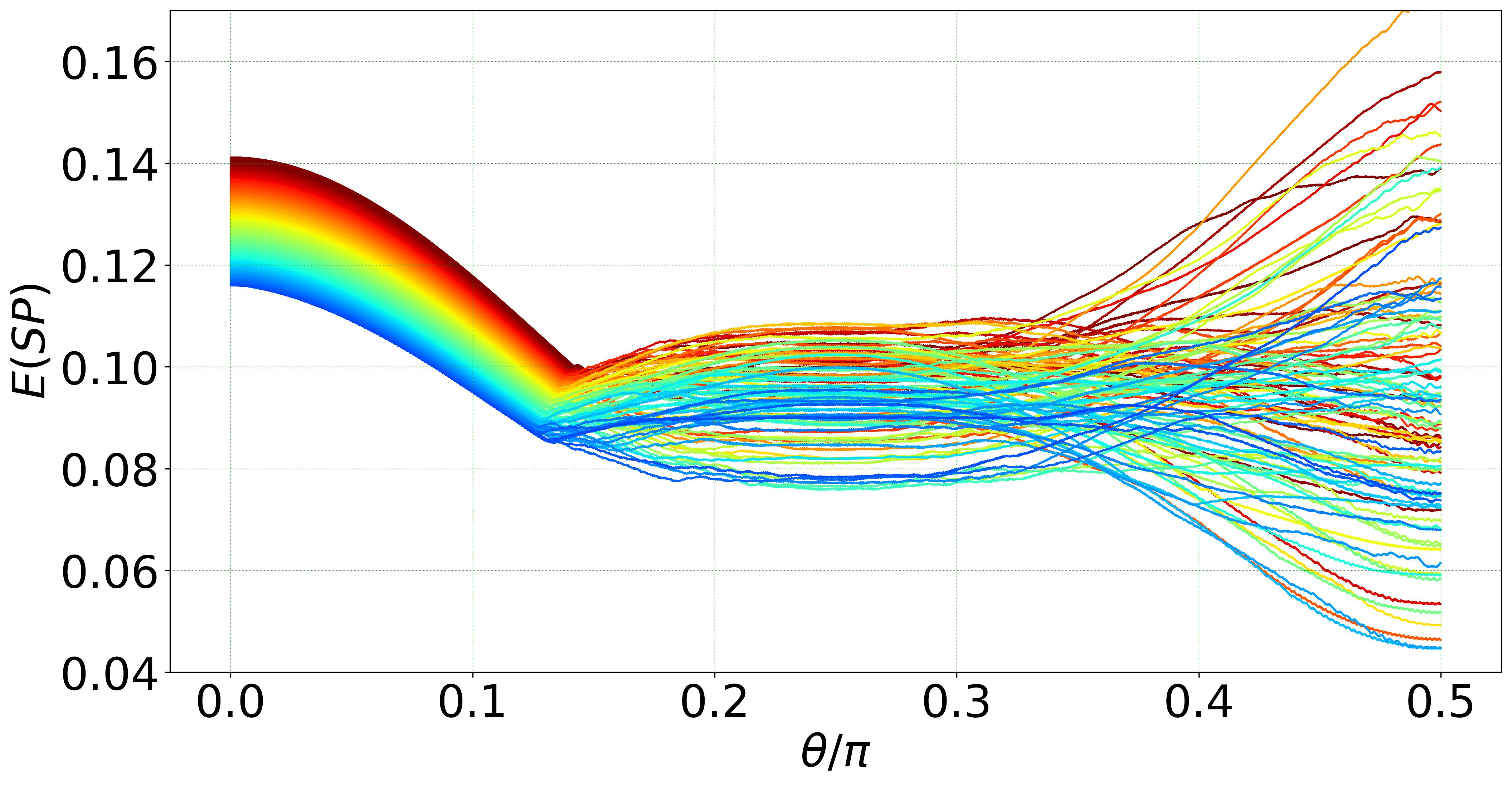}
\includegraphics[width=5cm]{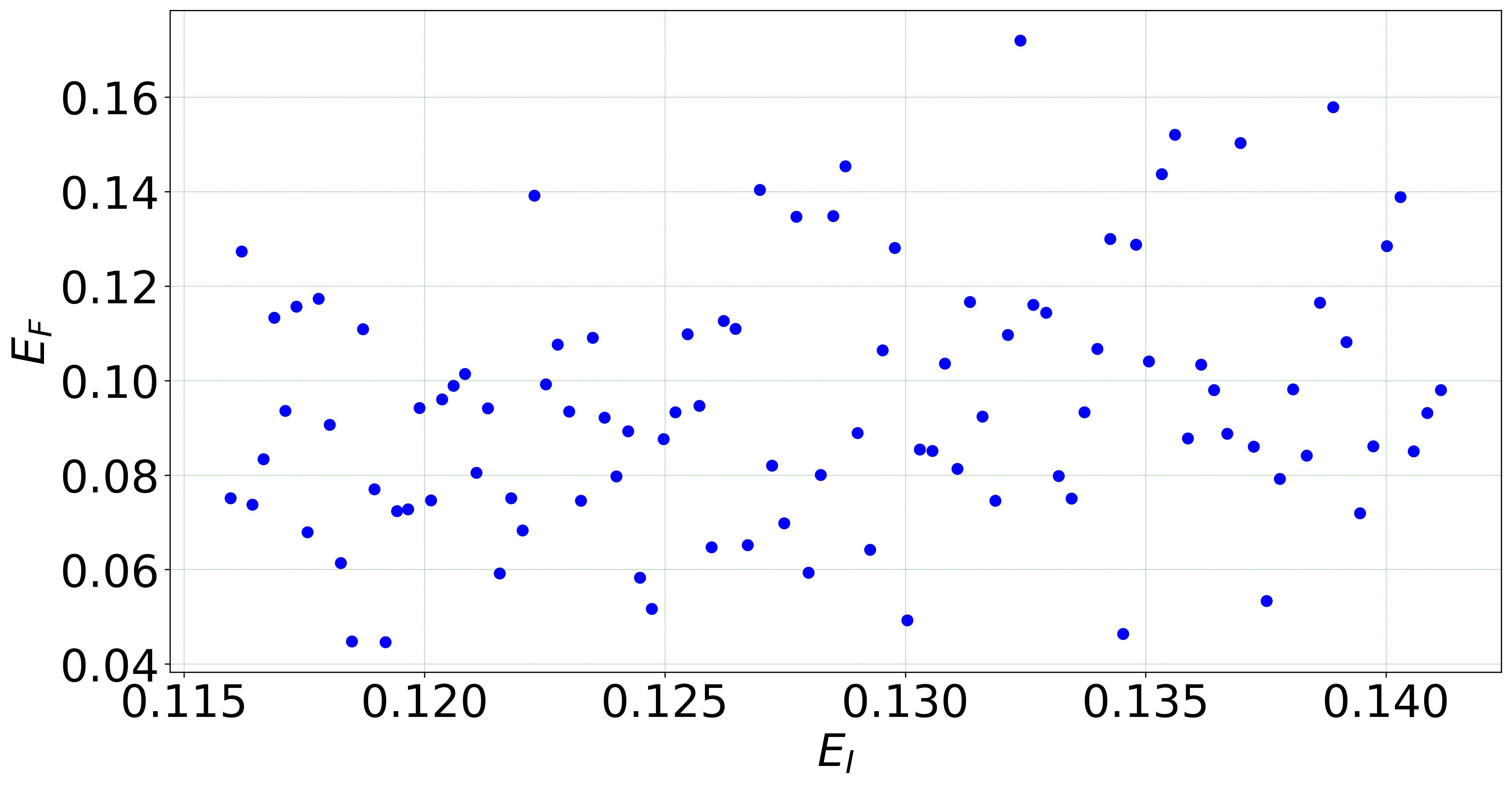}
\caption{{\textbf{Chaos in energy space.} 
Here the simulations are of BHC5 with $u=0.4$, but very similar results are obtained for BHC3. The left panel shows the time dependent energy for a cloud of trajectories. The right panels shows a scatter diagram of $E_{final}$ versus $E_{initial}$. The sweep is quasistatic with  $\dot{\theta} = \pi/2 \times 10^{-4}$.} 
}
\label{fig7_5}
\end{figure}

\clearpage
\sectA{Quantum vs semiclassical evolution}

Here we display additional panels for \Fig{f2}. 
The panel in \Fig{f2b} shows the BHC3 dynamics in the diabatic regime, where we can identify what we call longitudinal ``jumps" in energy. The panels in \Fig{f2c} shows the dynamics for BHC5, featuring breakdown of detailed QCC in the quasistatic regime, and longitudinal dynamics, as in BHC3, in the diabatic regime.

\ \\

\begin{figure}[tbh!]
\centering

\includegraphics[width=8cm]{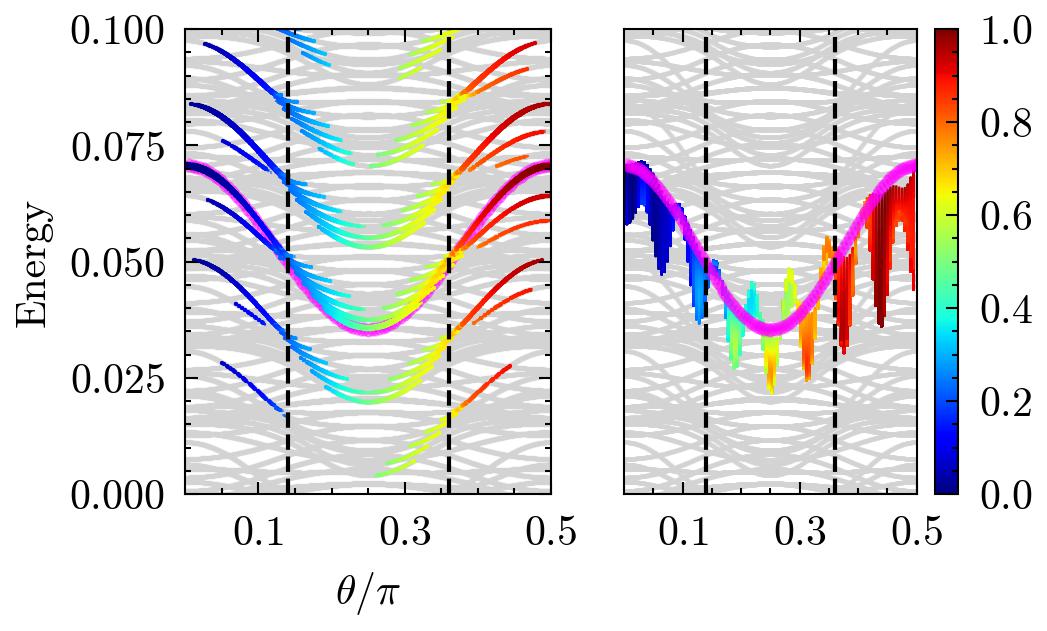}
\includegraphics[width=8cm]{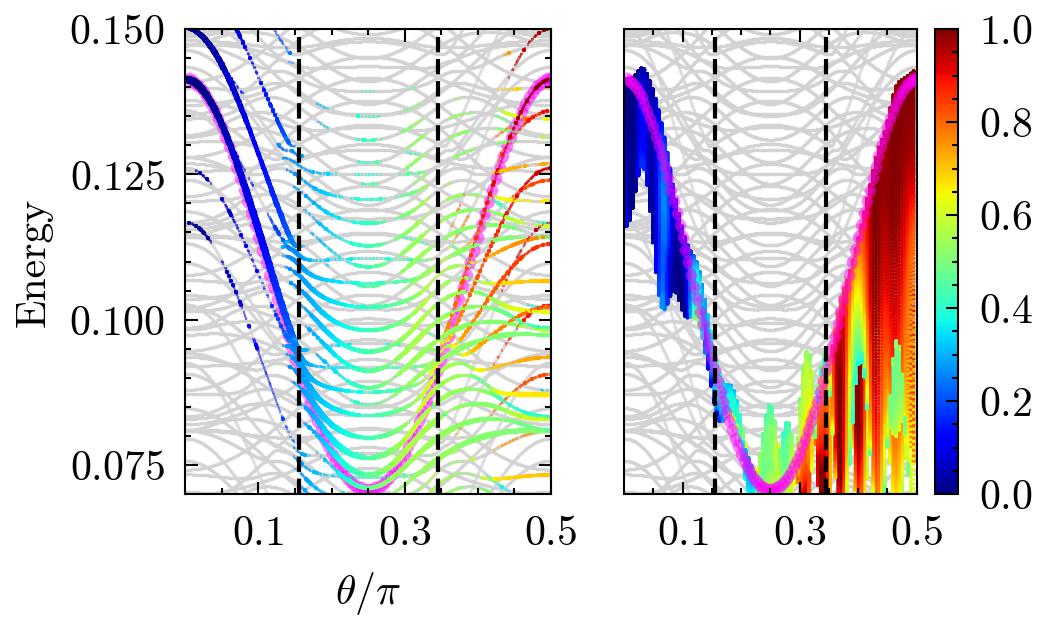}

\caption{\textbf{Quantum vs semiclassical evolution for BHC3.} 
See caption of \Fig{f2}. Here the simulations are for BHC3 in the diabatic regime 
The left panels are for $u=0.2$
with $\dot{\theta} = 4\pi/2 \times 10^{-2}$, 
while the right panels are for $u=0.4$.
with $\dot{\theta} = \pi/2 \times 10^{-2}$.
Vertical black dashed lines marks interval of $\theta$, where the SP is unstable.
Magenta dots denote SP energy.}
\label{f2b}
%
%
\ \\ \ \\ \ \\ 
%
\centering

\includegraphics[width=8cm]{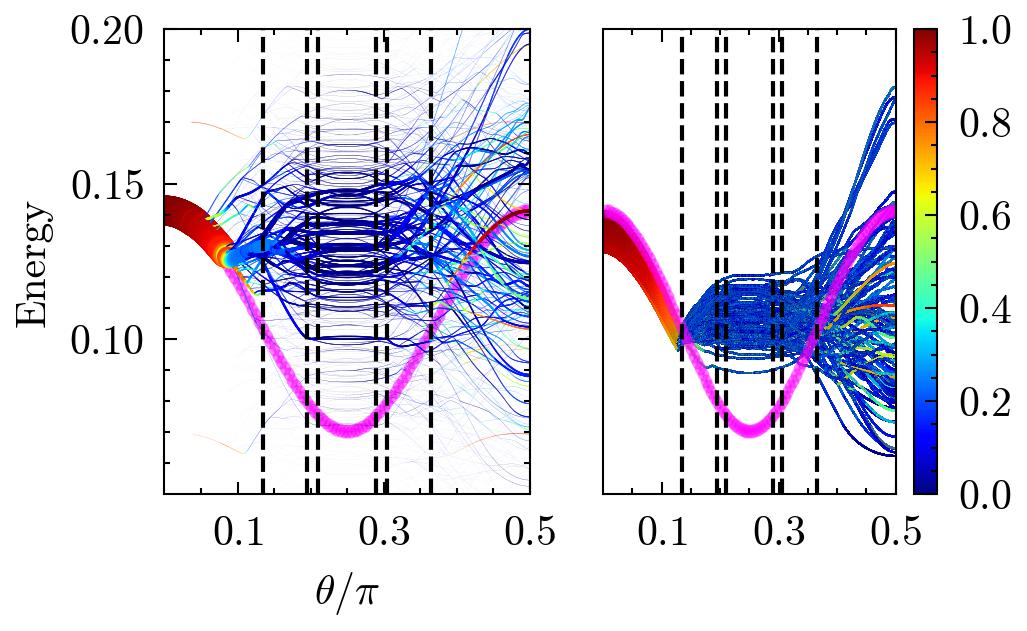}
\includegraphics[width=8cm]{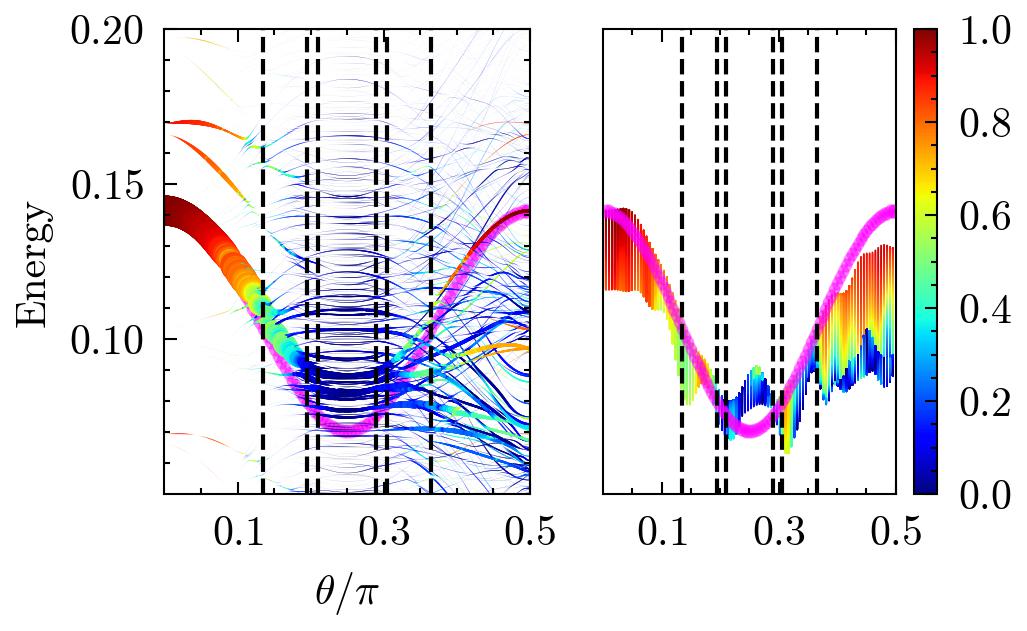}

\caption{\textbf{Quantum vs semiclassical evolution for BHC5.} 
See caption of \Fig{f2}. Here the simulations are for BHC5 with $u=0.4$.  
The left panels demonstrate the quasistatic regime ($\dot{\theta} = \pi/2 \times 10^{-5}$). 
The right panels are for the diabatic regime ($\dot{\theta} = \pi/2 \times 10^{-2}$). 
Vertical black dashed lines indicate where the SP becomes unstable based on Bogoliubov analysis.
}
\label{f2c}
\end{figure}

\clearpage
\section{Poincare sections}

\begin{figure}[tbh!]
\includegraphics[width=3.75cm, height=3.75cm]{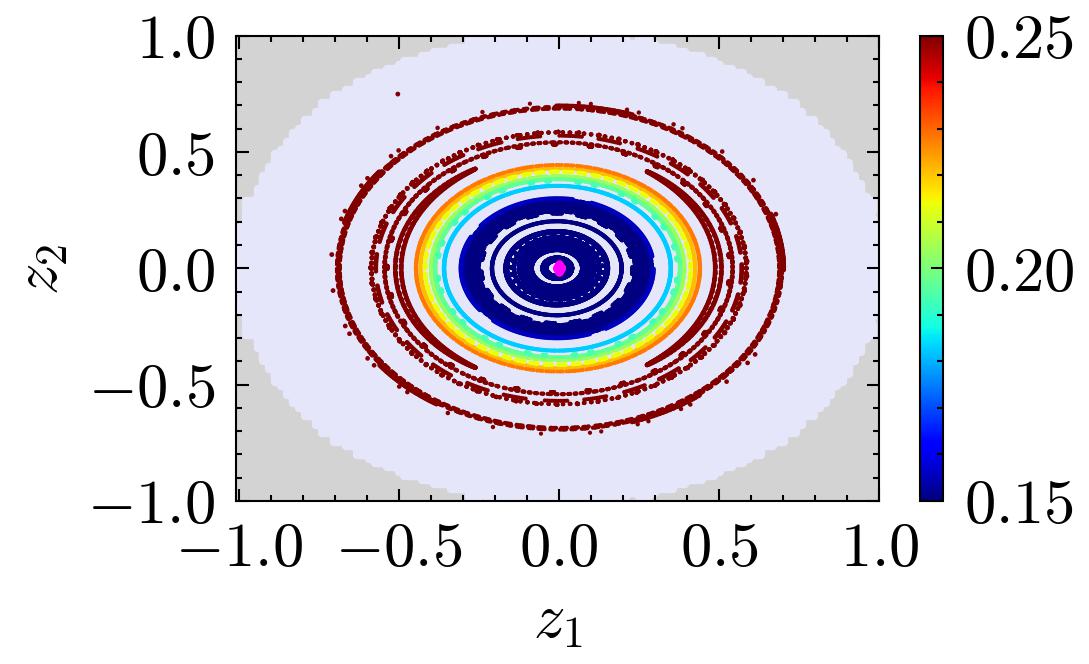}
\includegraphics[width=3.75cm, height=3.75cm]{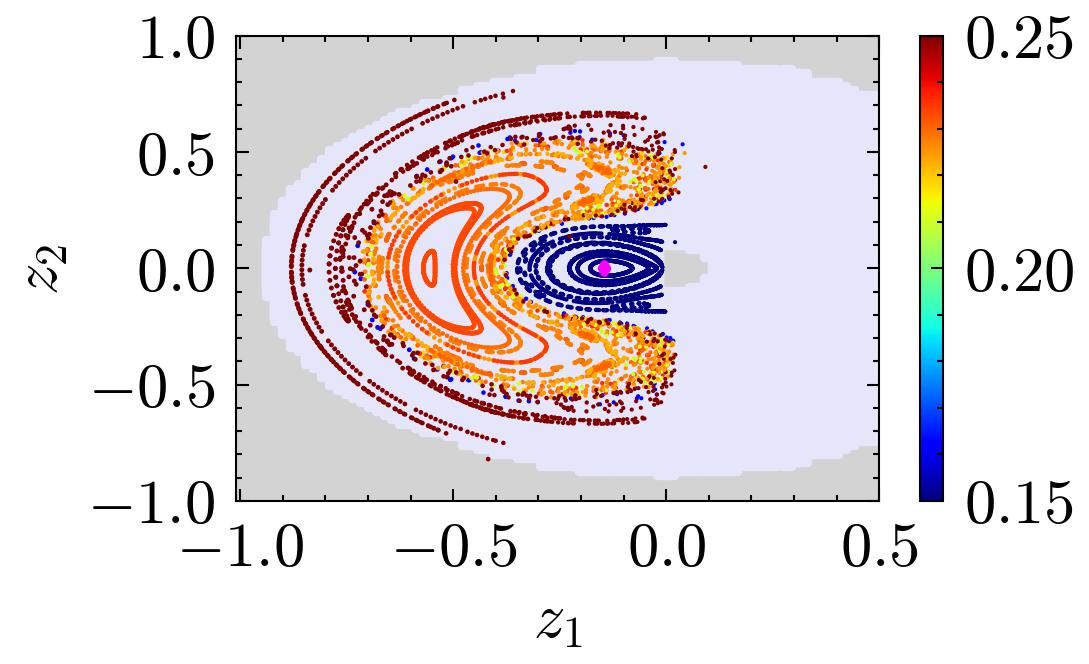}
\includegraphics[width=3.75cm, height=3.75cm]{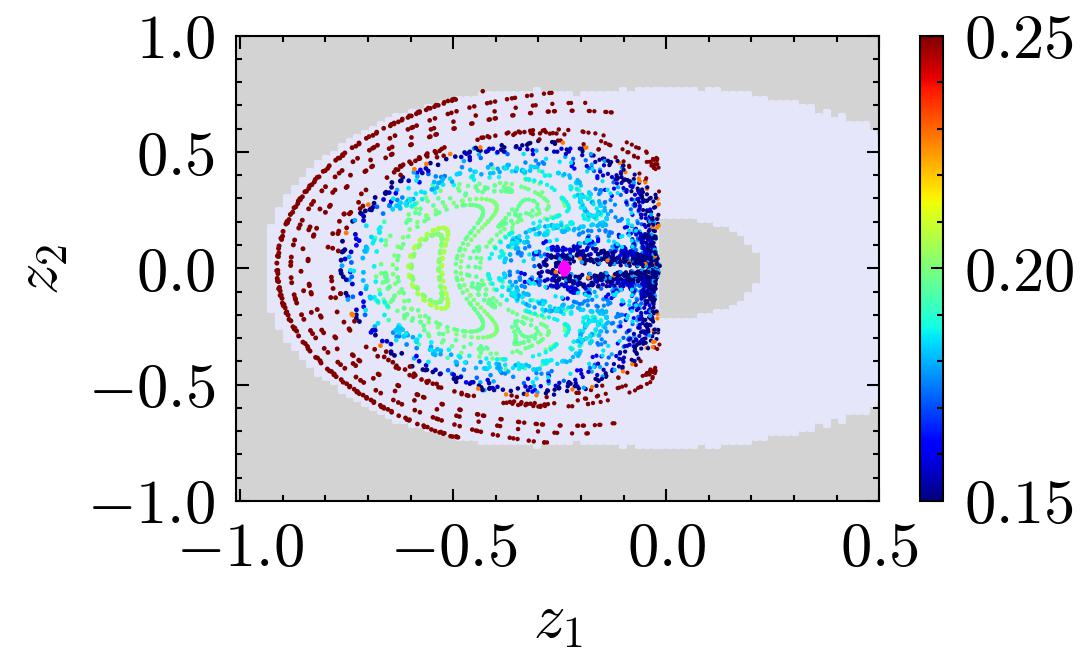}
\includegraphics[width=3.75cm, height=3.75cm]{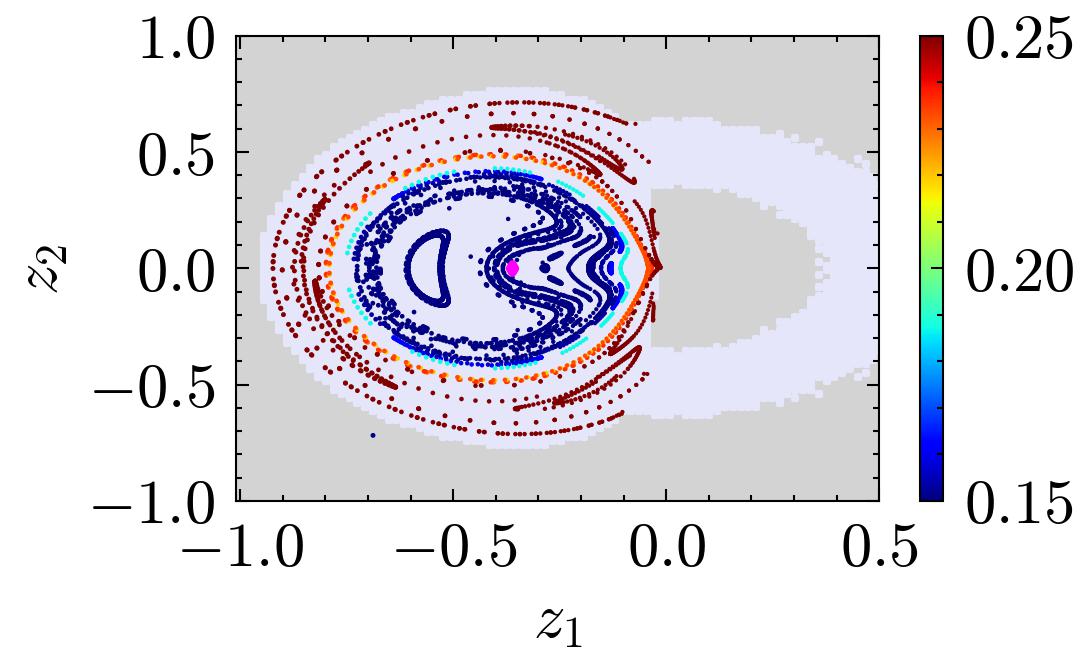}
\includegraphics[width=3.75cm, height=3.75cm]{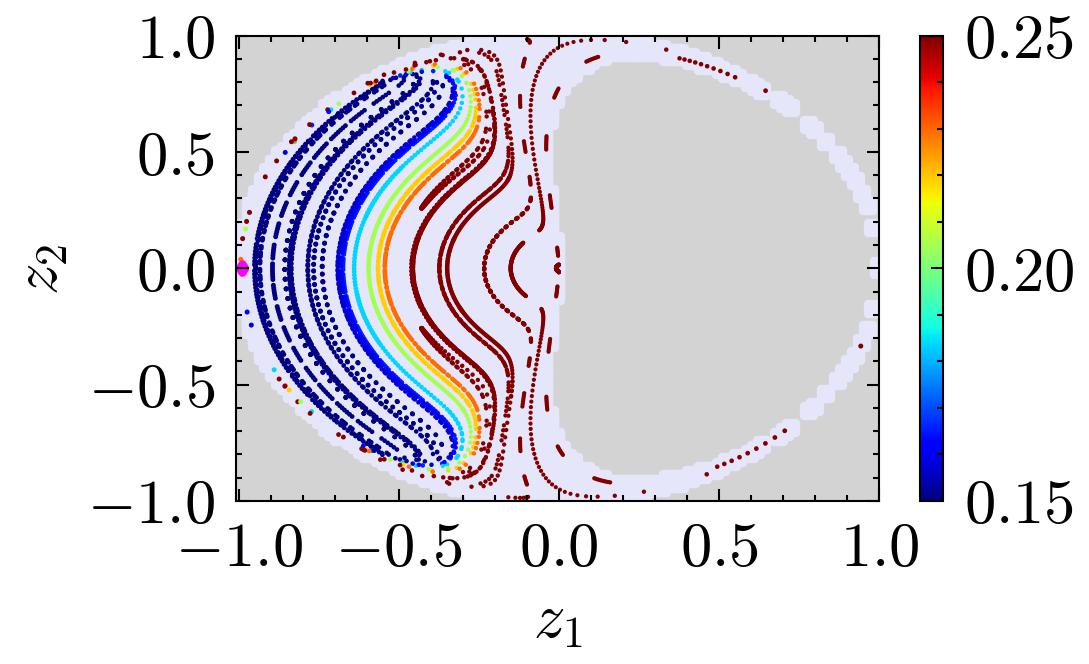}
\includegraphics[width=3.75cm, height=3.75cm]{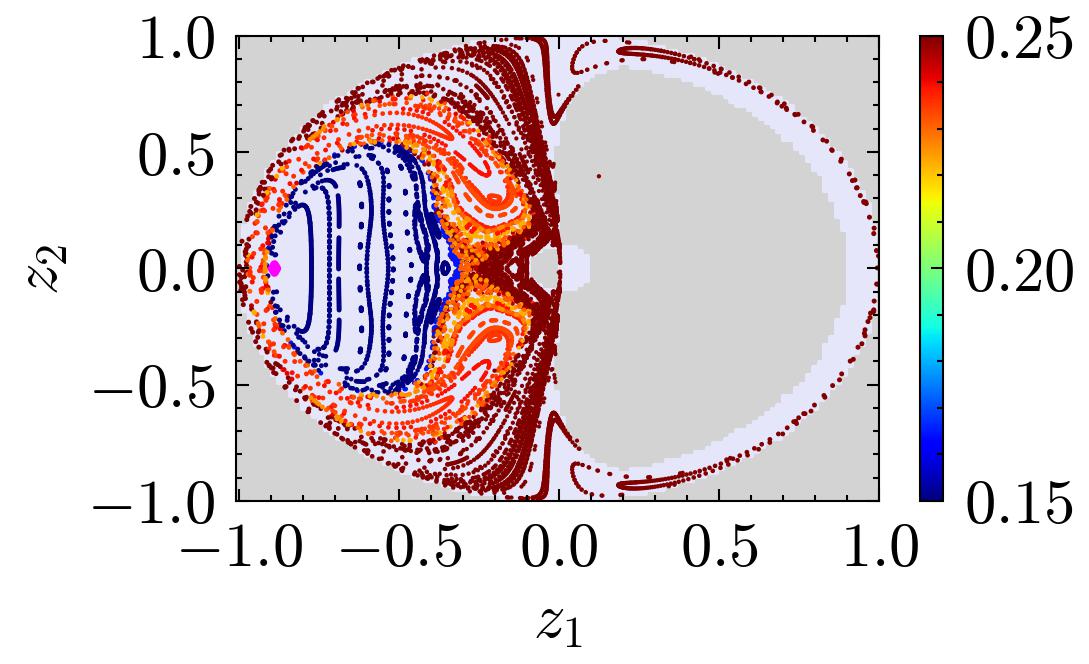}
\includegraphics[width=3.75cm, height=3.75cm]{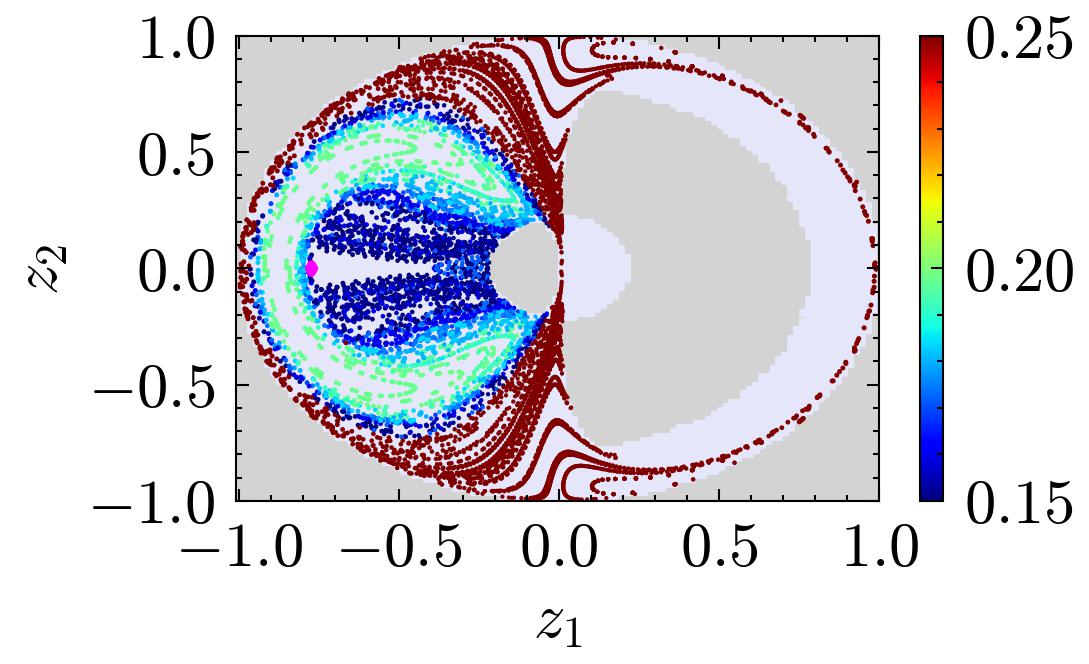}
\includegraphics[width=3.75cm, height=3.75cm]{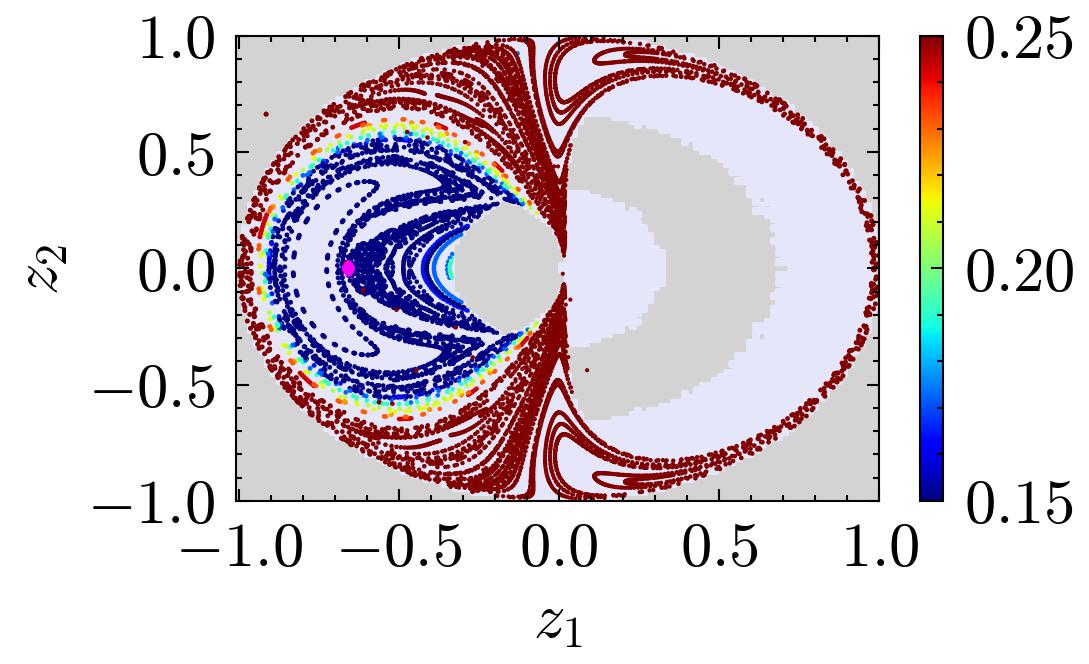}

\caption{{\bf Poincare sections.} BHC3 with ${u=0.4}$ and ${v=0.1}$. The upper row is for the first half of the sweep, while in the second row the dual images of the second half of the sweep are displayed. The color of a trajectory indicates the associated average value of $n_2(t)/N$. Magenta dot marks the position of the SP. The phasespace section cut is at $q_{2}=q_{2}(SP)$. 
In the first row from left to right $\theta= \{ 0.0157,0.33,0.471,0.628\}$.  
In the second row from right to left the dual panels are for $\theta= \{ 0.958,1.1,1.272,1.56\}$. 
Note that the SP is unstable in the range ${0.471<\theta<1.1}$.  }
\label{fig7_2}
\end{figure}

\Fig{fig7_2} shows how phasespace looks like for a sequence of $\theta$ values. Initially the central SP is stable in the center of a regular island (first two panels). Then it become unstable (3rd panel). In an adiabtic scenario the cloud would stretch along a torus that goes through the SP. Then the SP departs from this torus. The panels in the second row are dual to the panels of the first row. Eventually the SP becomes stable again. Ideally the cloud will end back at the SP. But due to the presence of a stochastic strip, this reversibility is spoiled. Namely, the spreading of the cloud through a strip (``thick torus"), rather than ``thin  torus", is like free expansion that cannot be reversed.      

\ \\

\ \\

\sectA{Evolution of the semiclassical cloud}

\begin{figure}[tbh!]

\includegraphics[height=4.5cm]{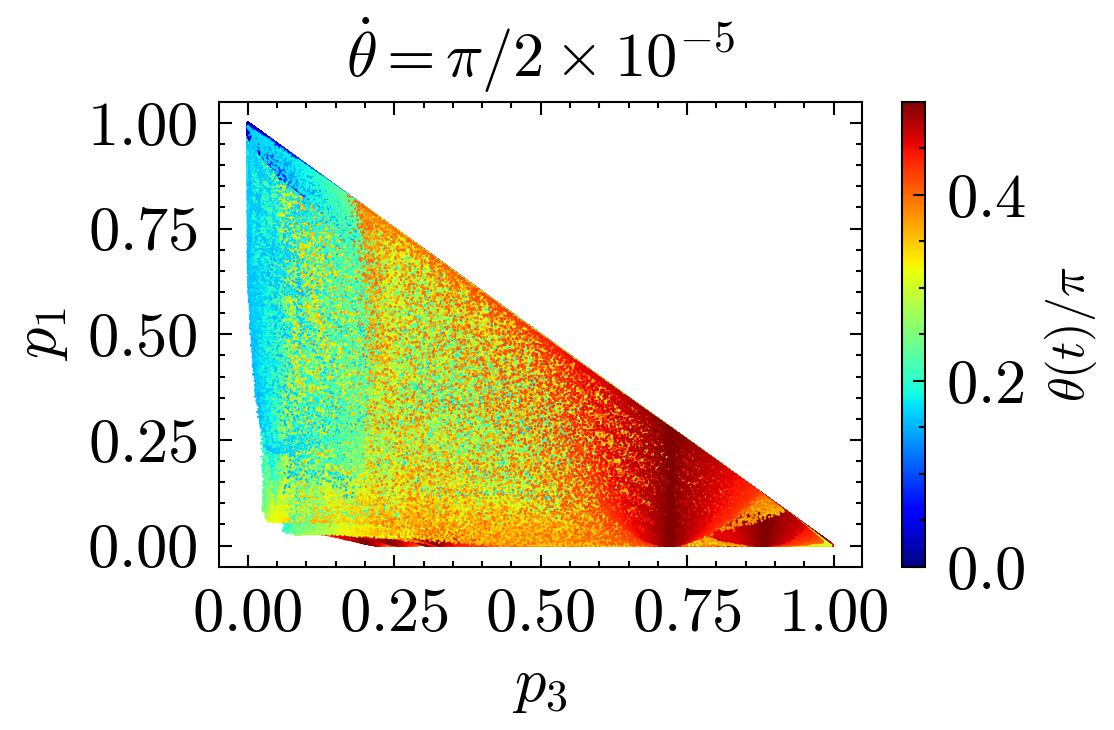}
\includegraphics[height=4.5cm]{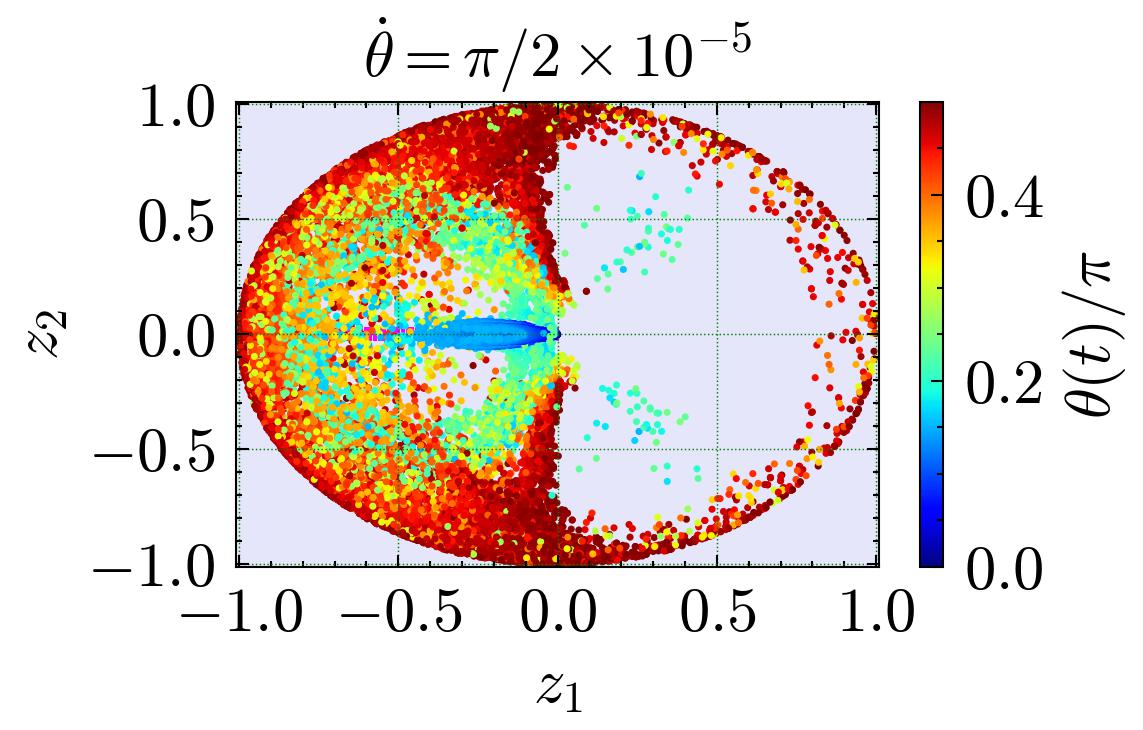}

\includegraphics[height=4.5cm]{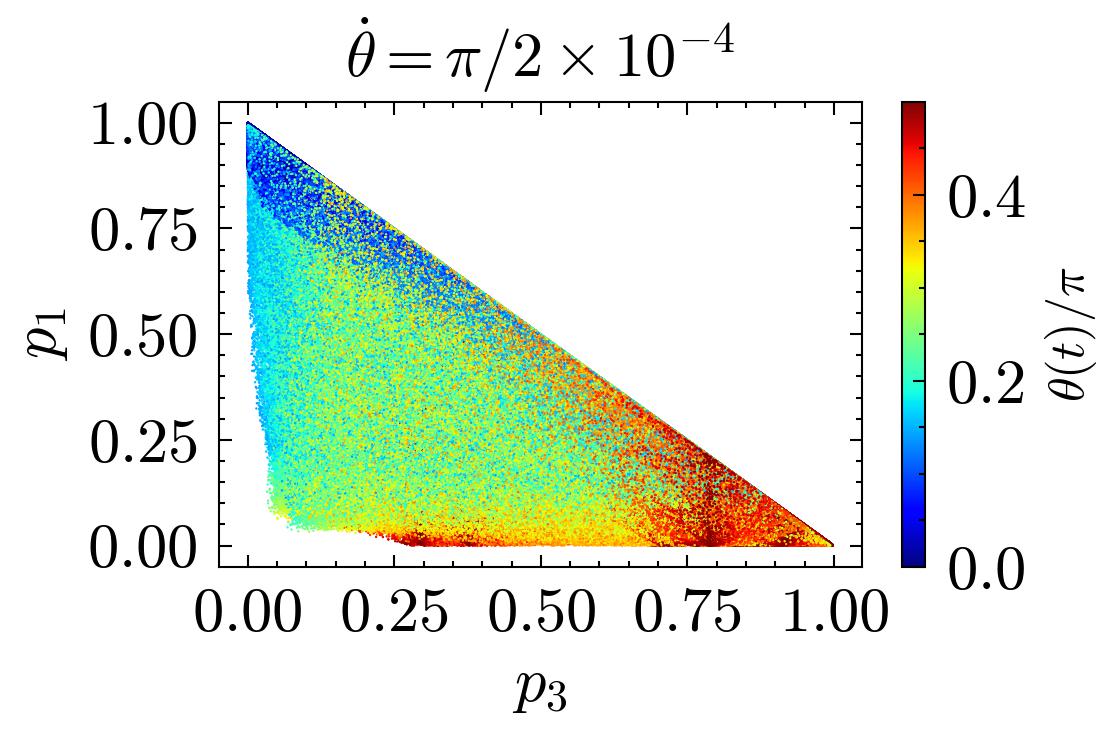}
\includegraphics[height=4.5cm]{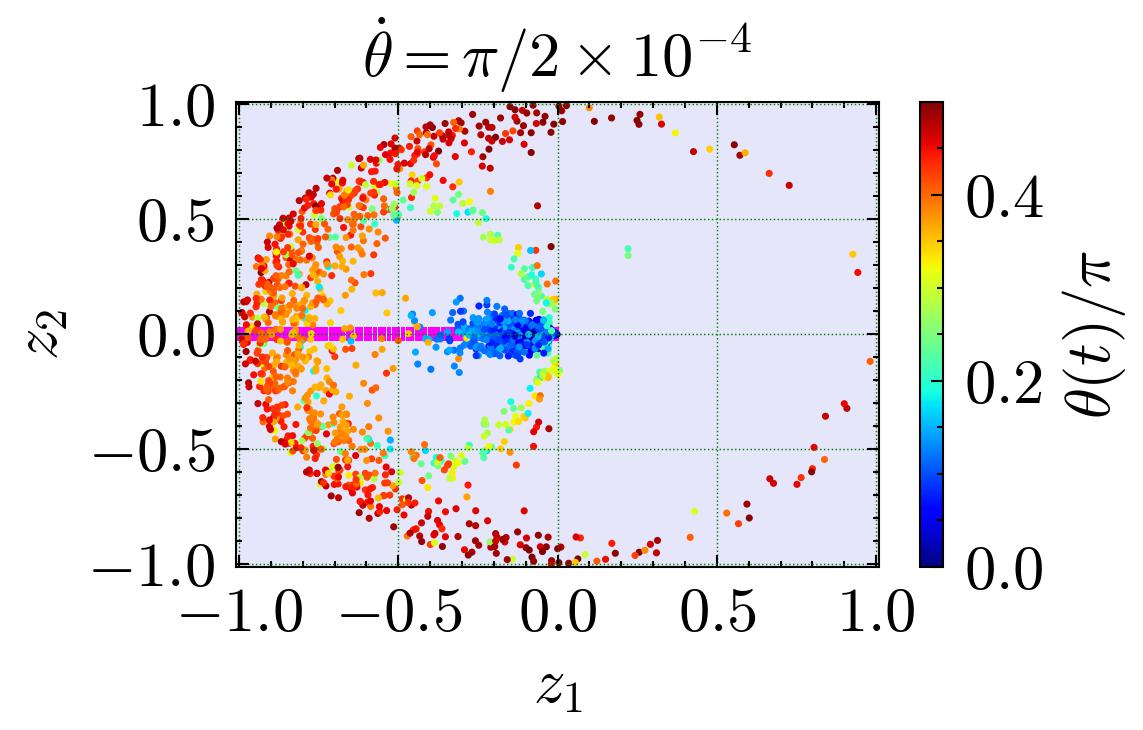}

\includegraphics[height=4.5cm]{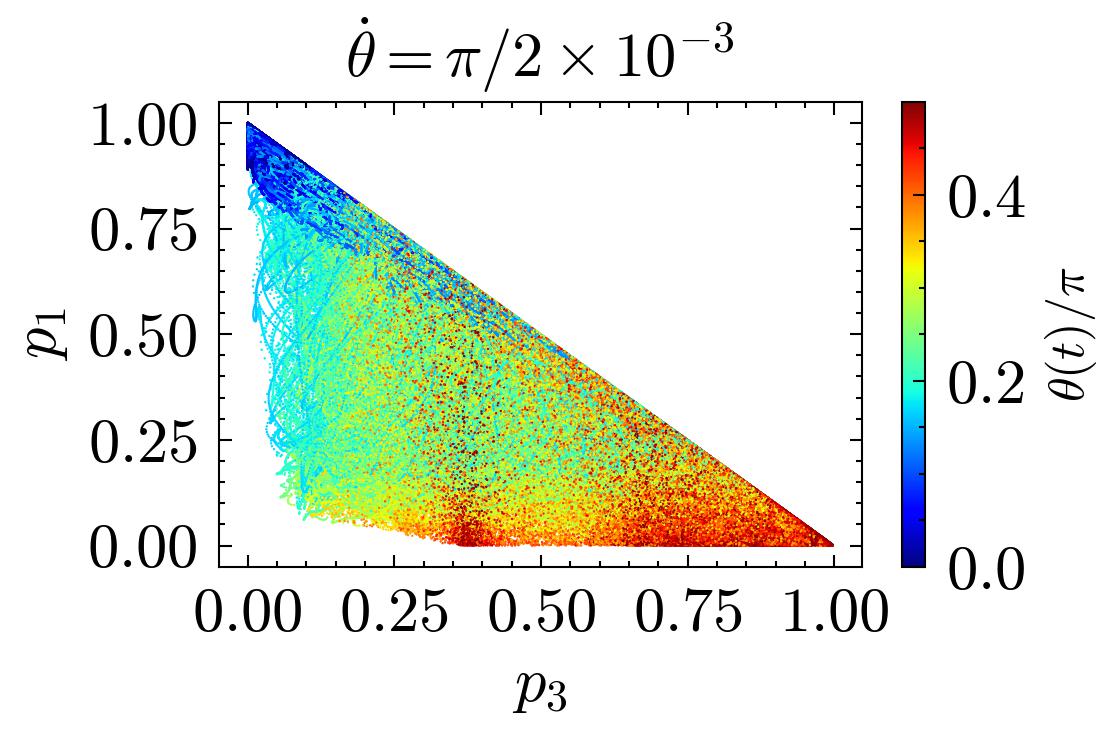}
\includegraphics[height=4.5cm]{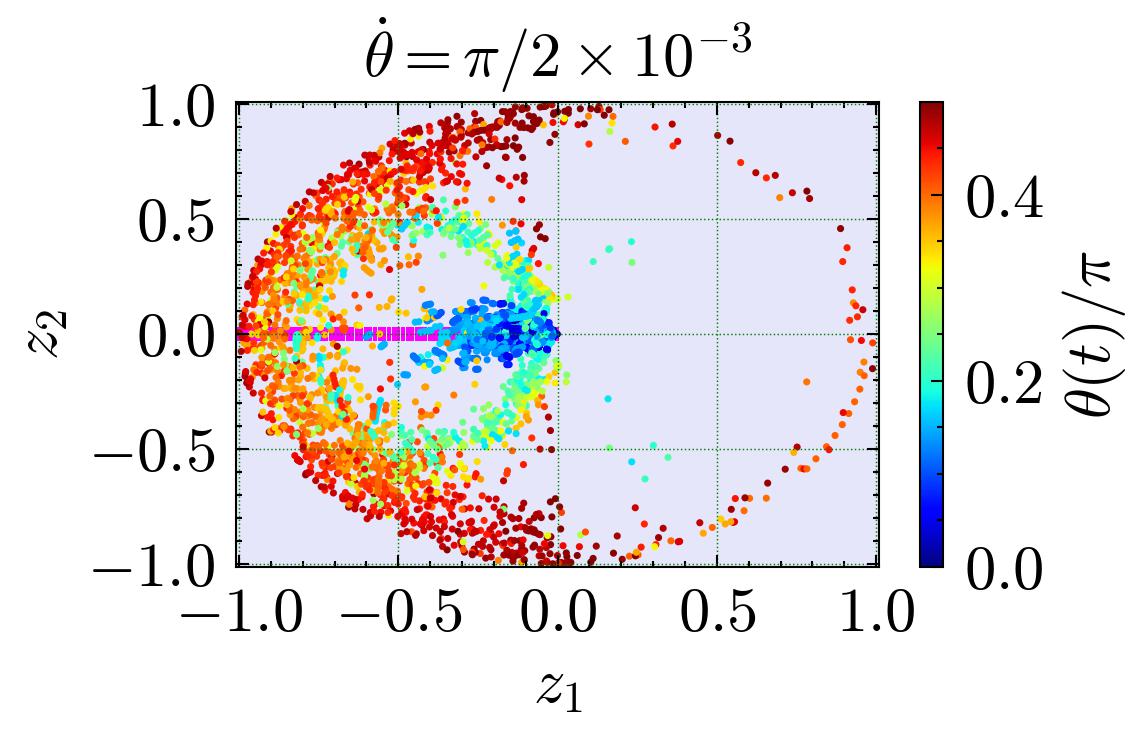}

\includegraphics[height=4.5cm]{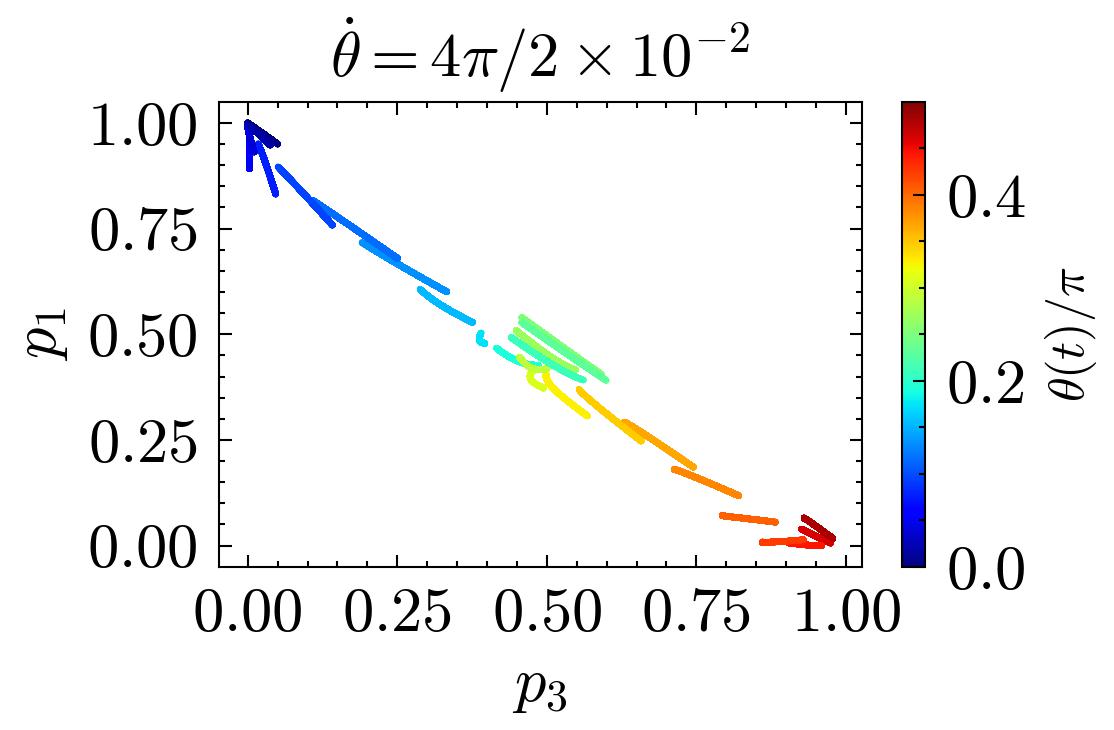} 
\includegraphics[height=4.5cm]{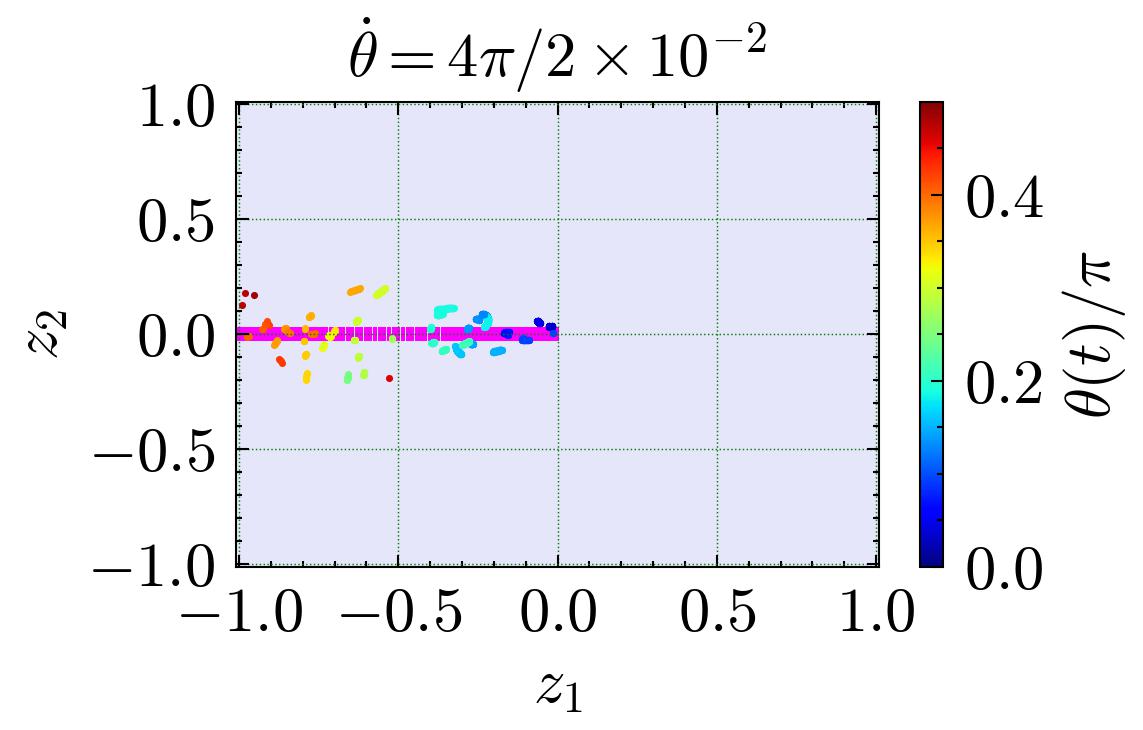}

\caption{\textbf{Evolution of the semiclassical cloud.} The evolving cloud of BHC3 for $u{=}0.4$ and $v{=}0.1$.
The left column provides images of the evolving cloud in occupation space, while the right columns are the corresponding phasespace sections. The initial cloud (blue) starts at ${p_1=1}$ which is the center (${z=0}$) of the phasespace plots. Subsequently, there are several snapshots. Each snapshot is colored by its $\theta(t)$. 
The phasespace cut is at $q_{2}=q_{2}(\text{SP}; \theta)$.}
\label{fig7_3}
\end{figure}

The evolution of the cloud in occupation space, and the corresponding phasespace sections are illustrated in \Fig{fig7_3}. In the diabtic regime (lower right panel) the cloud roughly follows the SP. For slower sweep, the cloud has the time to depart from the SP. It is stretched along a torus that evolves adiabatically to ${|z|=1}$. Note that ${|z|=1}$ implies that the 1st site is totally depleted, hence this circle should be regarded as a single point that represents a pole of a Bloch-like sphere. This pole is situated opposite to the $z{=}0$ origin.

\ \\

\clearpage
\end{document}